# Using Principal Progression Rate to Quantify and Compare Disease Progression in Comparative Studies

**Changyu Shen[1], Menglan Pang[1], Ling Zhu[1], Lu Tian[2]**

[1] Biogen, 225 Binney Street, Cambridge, MA 02142, USA

[2] Department of Biomedical Data Science, Stanford University School of Medicine, Palo Alto, CA 94305, USA

**Corresponding Author:** Changyu Shen, Biogen, 225 Binney Street, Cambridge, MA 02142, USA.

Email: changyu.shen@biogen.com

# ABSTRACT

In comparative studies of progressive diseases, such as randomized controlled trials (RCTs), the mean Change From Baseline (CFB) of a continuous outcome at a pre-specified follow-up time across subjects in the target population is a standard estimand used to summarize the overall disease progression. Despite its simplicity in interpretation, the mean CFB may not efficiently capture important features of the trajectory of the mean outcome relevant to the evaluation of the treatment effect of an intervention. Additionally, the estimation of the mean CFB does not use all longitudinal data points. To address these limitations, we propose a class of estimands called Principal Progression Rate (PPR). The PPR is a weighted average of local or instantaneous slope of the trajectory of the population mean during the follow-up. The flexibility of the weight function allows the PPR to cover a broad class of intuitive estimands, including the mean CFB, the slope of ordinary least-square fit to the trajectory, and the area under the curve. We showed that properly chosen PPRs can enhance statistical power over the mean CFB by amplifying the signal of treatment effect and/or improving estimation precision. We evaluated different versions of PPRs and the performance of their estimators through numerical studies. A real dataset was analyzed to demonstrate the advantage of using alternative PPR over the mean CFB.

**KEY WORDS:** change from baseline, estimation precision, effect size, principal progression rate, statistical power

## 1. INTRODUCTION

Change from baseline (CFB) of a continuous outcome has been a popular metric to quantify clinical, physiological, and molecular change related to evolvement of disease status. In many randomized control trials (RCTs), the mean CFB at a pre-specified post-baseline time point in the target population has been a standard estimand used to summarize the efficacy of the intervention of interest,[1] mostly due to its intuitive interpretation and easy estimation. Well established statistical tools such as Analysis of Covariance (ANCOVA) and Mixed-effects Models for Repeated Measures (MMRM) can improve the precision in estimating the mean CFB. Subsequently, statistical power can also be improved.[2-5]

The mean CFB as an estimand has several limitations. First, it focuses solely on "where" the outcome reaches without accounting for "how" it gets there. Therefore, it may not be the right estimand to adequately compare trajectories. For example, it is possible that trajectory A dominates trajectory B over the entire period of follow-up except at the end where the two are equal. In this case, it is debatable whether A and B have the same overall progression, despite the mean CFB at the end of follow-up suggests they are equal. Second, even if trajectories A and B differ at the end of the follow-up, the maximum difference between them may occur earlier, meaning the mean CFB misses the strongest signal. Third, the improvement in power by ANCOVA can be limited if there is no available baseline covariate with a decent correlation with the outcome at the chosen time point.

For progressive diseases such as Alzheimer's and Parkinson's diseases, the trajectory of the population mean of relevant clinical outcome assessments (COAs) is typically monotone, e.g., the mean COA either remains stable or becomes worse. Raket proposed a class of non-linear mixed-effects models with specific model parameters to capture treatment effects measured from various perspectives.[6] Nonetheless, this approach requires correct assumption about the features of the longitudinal treatment effect. In this article, we propose a new class of estimands called Principal Progression Rate (PPR) as a measure of the speed of disease progression for a population. It covers a broad spectrum of estimands, including the mean CFB. Essentially, the PPR is a weighted average of local or instantaneous slope (e.g. derivative) of the mean trajectory of the outcome across the entire follow-up period. This new framework offers three advantages. First, it allows for flexible specification of estimands through the weight function, enabling emphasis or deemphasis of certain follow-up period. Second, certain PPRs in this class offer augmentation of signal of treatment effect and/or reduction of estimation error, potentially leading to substantial improvement in statistical power. Third, a PPR can be easily estimated non-parametrically with panel longitudinal data using standard statistical software. These features make the PPR a useful estimand to be used in RCTs and other comparative studies.

In the following sections, we provide a detailed explanation on ideas leading to the definition and estimation of PPR in Section 2. In Section 3, we present numerical examples demonstrating statistical properties of the PPR estimands and their estimators. We apply our method to a real dataset in Section 4 and conclude this article with a discussion in Section 5.

## 2. METHODS

### 2.1 Summary of disease progression

Let $f(t)$ (where $0 \leq t \leq 1$) be a smooth trajectory of the population mean of a continuous outcome, with higher values indicating more disease severity, over a follow-up period indexed by $t$ (standardized to have length 1). Since we focus on progressive disease in this article, $f(t)$ is a non-decreasing function of time. In a typical RCT, the estimand of an efficacy endpoint is usually the mean CFB at the end of the follow-up period, defined as:

$$r_1(f) = f(1) - f(0).$$

Therefore, $r_1(f)$ can be viewed as a slope that characterizes an overall progression rate.

Although $r_1(f)$ enjoys conceptual simplicity, it only characterizes the magnitude of progression at the beginning and end of the follow-up period, ignoring other features of the curve $f(t)$. For instance, local "slope" could carry valuable information about disease progression. A natural extension of $r_1(f)$ is to consider more than two time points. Let $0 = t_1 < t_2 < \cdots < t_{m-1} < t_m = 1$ be a set of $m$ time points. Just as $r_1(f)$ can be viewed as the slope of a fitted line connecting two points, $(0, f(0))$ and $(1, f(1))$, we can also calculate the slope of a line fitted to the $m$ time points using the ordinary least-square (OLS) method

$$\frac{\sum_{i=1}^m (t_i - \bar{t}) f(t_i)}{\sum_{i=1}^m (t_i - \bar{t})^2}, \bar{t} = \frac{1}{m} \sum_{i=1}^m t_i.$$

We can rewrite the slope as

$$r_{L,t}(f) = \sum_{i=2}^m L_i \frac{f(t_i) - f(t_{i-1})}{t_i - t_{i-1}}, L_i = \frac{(t_i - t_{i-1}) \sum_{j=i}^m (t_j - \bar{t})}{\sum_{k=1}^m (t_k - \bar{t})^2} > 0,$$

$$\sum_{i=2}^m L_i = 1.$$

Here the letter "$L$" refers to the weight vector corresponding to the OLS linear fit, and "$t$" denotes the vector of time points. Therefore, $r_{L,t}(f)$ can be viewed as a weighted average of local slopes of all consecutive pairs of time points. It is clear that $r_{L,t}(f) \geq 0$ as $f(t)$ is non-decreasing. In the special case where the $m$ time points are equally spaced, with $t_1 = 0$ and $t_m = 1$, the weight $L_i$ is given by:

$$L_i = \frac{6(i-1)(m+1-i)}{m(m-1)(m+1)}, i = 2, \ldots, m.$$

It is clear that the pair of time points in the "middle" of the follow-up has higher weight in this case.

An alternative interpretation of $r_1(f)$ is that it is roughly twice the area between $f(t)$ and the horizontal line $y = f(0)$, if $f(t)$ is approximately linear in $t$. Following this interpretation, we can extend the definition to the $m$ equally spaced time points with $t_1 = 0$ and $t_m = 1$. Specifically, twice the area between the curve and the horizontal line $y = f(0)$ based on the rectangular rule can be approximated by

$$\frac{2}{m-1}\sum_{i=2}^{m}[f(t_i) - f(t_1)] \approx \frac{2}{m}\sum_{i=2}^{m}[f(t_i) - f(t_1)]$$

The latter quantity in the expression above can be rewritten to define a summary of progression as

$$r_{A,t}(f) = \sum_{i=2}^{m} A_i \frac{f(t_i) - f(t_{i-1})}{t_i - t_{i-1}}, \quad A_i = \frac{2(m-i+1)}{m(m-1)} > 0,$$

$$\sum_{i=2}^{m} A_i = 1.$$

Here the letter "$A$" denotes a vector of weight corresponding to the area under the curve principle. Similar to $r_{L,t}(f)$, $r_{A,t}(f)$ is also a weighted average of slopes of all consecutive pairs of time points. However, unlike $r_{L,t}(f)$, $r_{A,t}(f)$ assigns more weight to the earlier time points and less weight toward the end of the follow-up period.

## 2.2 Principal Progression Rate

In general, we can define a discrete Principal Progression Rate (PPR) as

$$r_{w,t}(f) = \sum_{i=2}^{m} w_i \frac{f(t_i) - f(t_{i-1})}{t_i - t_{i-1}}, \quad \sum_{i=2}^{m} w_i = 1, w_i \geq 0.$$

Clearly, $r_1, r_{L,t}()$ and $r_{A,t}()$ are all special cases of the class of discrete PPRs. By definition, $r_{w,t}(f) \geq 0$, and as a measure of overall disease progression, it offers a major advantage of allowing flexible weighting across different periods of follow-up time. We can rewrite $r_{w,t}(f)$ as

$$r_{w,t}(f) = \sum_{i=1}^{m} v_i f(t_i), v_i = \begin{cases} -\dfrac{w_{i+1}}{t_{i+1} - t_i}, & i = 1 \\ \dfrac{w_i}{t_i - t_{i-1}} - \dfrac{w_{i+1}}{t_{i+1} - t_i}, & 1 < i < m. \\ \dfrac{w_m}{t_m - t_{m-1}}, & i = m \end{cases}$$

It is easy to see that $\sum_{i=1}^{m} v_i = 0$. Therefore, the PPR is a scaled contrast.

The continuous version of PPR (e.g., $m \to \infty$) can be defined as

$$r_w(f) = \int_0^1 w(t) f'(t) dt, \quad \int_0^1 w(t) dt = 1, \ w(t) \geq 0.$$

Since the local slope is proportional to the integration of derivatives, any $r_{w,t}(f)$ with $\mathbf{w} = (w_2, w_3, \ldots w_m)^T$ can be written as $r_{w_c}(f)$, where $w_c$ is a step function defined as: $w_c(t) = w_i/(t_i - t_{i-1})$, $t_{i-1} < t \leq t_i$ ($i = 2, \ldots, m$). It is easy to see that $r_1(f)$ can be viewed as a continuous PPR with $w(t) \equiv 1$, or $Beta(1,1)$, where $Beta(\alpha, \beta)$ corresponds to a Beta probability density function with parameters $\alpha$ and $\beta$. As $m \to \infty$ with $m$ equally spaced time points, $r_{L,t}(f)$ converges to $r_L(f)$ with $L(t) = Beta(2,2) = 6t(1-t)$. In other words, $r_L(f)$ is the slope of an OLS linear fit to the entire curve of $f(t)$. Similarly, the continuous version of $r_{A,t}(f)$ is $r_A(f)$ with $A(t) = Beta(1,2) = 2(1-t)$. Again, we see that $r_L(f)$ assigns more weight to the middle of the follow-up window whereas $r_A(f)$ assigns more weight to the beginning. In **Figure 1**, we illustrate the weight functions of several PPRs discussed above.

The PPR offers a general framework for defining a summary measure of disease progression by allowing progression at different time points to be weighted differently. If progression at the beginning, middle or the end of follow-up is more/less important, the corresponding period can be given greater/smaller weight. For example, in RCTs of Alzheimer's diseases, progression at the beginning of the follow-up is often less critical since it typically takes time for a medical intervention to have impact on the COAs. Therefore, a smaller weight for the early follow-up period would be more appropriate compared to the mean CFB, which assigns equal weight to each time point. Furthermore, a properly chosen weight function can substantially improve estimation precision over the mean CFB. In Sections 2.3 and 2.4, we will discuss the following two mechanisms by which the use of PPR can enhancement of statistical power:

1) **Signal amplification:** If more weight is assigned to the time period where treatment effect on progression speed is the most, then the signal of treatment effect will be amplified.

2) **Precision improvement:** For certain type of weight, the estimation has higher precision than the mean CFB.

### 2.3 Signal of Treatment Effect

Let $f(t)$ and $h(t)$ represent the mean outcome trajectory for the control and treatment arms of an RCT, respectively. Let $\Delta(t) = f(t) - h(t)$ be the difference between the two trajectories at time $t$. We consider treatment effect as measured by the difference in continuous version of PPR. The same principle applies to the discrete version of PPR. We consider the following measure of treatment effect:

$$\Delta_w = r_w(f) - r_w(h) = \int_0^1 w(t)\Delta'(t)dt = w(1)\Delta(1) - w(0)\Delta(0) - \int_0^1 w'(t)\Delta(t)dt,$$

where $\Delta'(t)$ is the difference in progression speed between the two arms at time $t$. Note that in an RCT, $\Delta(0) = 0$, but we keep this term in the equation above for purpose that will be described in Section 2.4. To simplify the interpretation of $\Delta_w$, we consider the following parameter space throughout this article:

$$\textit{Null}: \Delta(t) \equiv 0, 0 \leq t \leq 1$$

*Alternative 1 (treatment benefit)*: $\Delta(t) \geq 0$ for $0 \leq t < 1$ and $\Delta(1) > 0$

*Alternative 2 (treatment harm)*: $\Delta(t) \leq 0$ for $0 \leq t < 1$ and $\Delta(1) < 0$

In other words, we consider the two trajectories to be either entirely identical or one dominates the other with at least some separation at the end of the follow-up. It is easy to see that the null corresponds to $\Delta_w = 0$. To better understand the behavior of $\Delta_w$ under the two alternatives, we introduce the concept of *consistency* of $\Delta_w$. By consistency of $\Delta_w$, we mean that under Alternatives 1 (treatment benefit), $\Delta_w > 0$, and under Alternative 2 (treatment harm), $\Delta_w < 0$. It is easy to see that without any further restriction on $\Delta(t)$, we need $w'(t) < 0$ to maintain consistency. In other words, to ensure consistency between the sign of $\Delta_w$ and the corresponding Alternatives, $w(t)$ should be a decreasing function.

The key idea of signal amplification in treatment effect roots in the principle of finding a $w(t)$ that is "correlated" with $\Delta'(t)$. This concept can be formally expressed as:

$$\Delta_w = \int_0^1 w(t)\Delta'(t)dt = Cov[w(T), \Delta'(T)] + \Delta_1, T \sim Uniform(0,1),$$

where $\Delta_1$ is the difference in mean CFB between the two groups. Therefore, if $w(T)$ and $\Delta'(T)$ are positively correlated where $T$ is uniform over $(0, 1)$, then $\Delta_w > \Delta_1$. In practice, the exact shape of $\Delta'(t)$ is unknown. However, we can rely on certain features of $\Delta'(t)$ to make an appropriate choice of $w(t)$ that is likely to match those features. Often these features put further restriction on the parameter space of Alternatives 1 and 2 such that we can still maintain consistency with a weight function that is not strictly decreasing. For instance, if $\Delta'(t)$ is known to be non-negative or non-positive over the interval $[0, 1]$, then any weight function $w(t) > 0$ will guarantee consistency of $\Delta(t)$. Another example is when the treatment effect on progression speed is constant, $\Delta'(t) = c \geq 0$ or $\Delta(t) = ct$, then $\Delta_w \equiv c$ regardless of the choice of $w(t)$. Hence consistency is maintained with any $w(t)$.

If treatment starts to reduce the progression speed immediately after the baseline and the effect gradually attenuates, resulting in a decreasing $\Delta'(t)$, then using $\Delta_A$ can amplify signal. The unique benefit of a decreasing weight function is that consistency is maintained even if additional assumptions about $\Delta'(t)$ are incorrect. On the other hand, it is possible that a therapeutic agent may not initially reduce progression speed but begins to slow progression at a later stage, and then this effect gradually attenuates. In this case, $\Delta'(t)$ first increases and then decreases. The corresponding behavior of $\Delta(t)$ is that it remains 0 for a period of time and then starts to increase followed by a plateau or decrease. As such, $\Delta_L$ can potentially amplify the signal relative to $\Delta_1$. To understand what extra restriction is required to maintain consistency when using $\Delta_L$, we need to have $\Delta_L > 0$ under Alternative 1 (or $\Delta_L < 0$ under Alternative 2). This requires:

$$\int_0^1 (t-0.5)\Delta(t)dt > 0 \Leftrightarrow \int_{0.5}^1 (t-0.5)\Delta(t)dt > \int_0^{0.5}(0.5-t)\Delta(t)dt.$$

Thus, to ensure consistency when using $\Delta_L$, the "average" of $\Delta(t)$ during the second half of the follow-up period must be higher than the "average" of $\Delta(t)$ during the first half. Here "average" is defined with respect to the probability density function $4|t-0.5|I(0 \le t \le 1)$.

### 2.4 Estimation of Difference in PPR

#### 2.4.1 Discrete PPR

We consider estimating treatment effect using panel longitudinal data. The estimand of treatment effect in terms of PPR can be defined as:

$$\Delta_{w,t} = r_{w,t}(f) - r_{w,t}(h) = \sum_{i=1}^m v_i \Delta(t_i).$$

Well-established approaches, such as Mixed-effects Models for Repeated Measures (MMRM), can be used to obtain the estimate $\hat{\Delta}(t_i)$ and the variance-covariance estimate $\hat{\sigma}_{ij}$ for $\hat{\Delta}(t_i)$ and $\hat{\Delta}(t_j)$. We have

$$\hat{\Delta}_{w,t} = \sum_{i=1}^{m} v_i \hat{\Delta}(t_i), \widehat{Var}[\hat{\Delta}_{w,t}] = \sum_{i=1}^{m} \sum_{j=1}^{m} v_i v_j \hat{\sigma}_{ij}.$$

Notably, since $\Delta(t_1) = \Delta(0) = 0$ in an RCT, any choice of $v_1$ leads to an unbiased estimate of $\Delta_{w,t}$. Thus, we can choose a $v_1^*$ that can reduce the variance without introducing any bias. Specifically, as shown in the **Supplementary Material S1,** we can construct a $\tilde{\Delta}_{w,t}$ that is the same as $\hat{\Delta}_{w,t}$ except that $v_1$ is replaced by

$$v_1^* = -\sum_{i=2}^{m} v_i \frac{\hat{\sigma}_{i1}}{\hat{\sigma}_{11}}.$$

Therefore, we have

$$\tilde{\Delta}_{w,t} = \sum_{i=2}^{m} v_i \left\{ \hat{\Delta}(t_i) - \frac{\hat{\sigma}_{i1}}{\hat{\sigma}_{11}} \hat{\Delta}(t_1) \right\}$$

with variance estimated as:

$$\widehat{Var}(\tilde{\Delta}_{w,t}) = \sum_{i=2}^{m} \sum_{j=2}^{m} v_i v_j \left[ \hat{\sigma}_{ij} - \frac{\hat{\sigma}_{i1}\hat{\sigma}_{j1}}{\hat{\sigma}_{11}} \right].$$

### 2.4.2 Continuous PPR

Standard parametric, semi-parametric and non-parametric (if data collection is frequent enough) method can be used to estimate $\Delta(t)$ and plug in the definition of $\Delta_w$ for estimation. We propose an approach based on the approximation of the integral $\int_0^1 w'(t)\Delta(t)dt$ through Gaussian Quadrature. Specifically,

$$\Delta_w \cong w(1)\Delta(1) - w(0)\Delta(0) - 0.5 \sum_{i=2}^{m-1} a_i w'(t_i)\Delta(t_i), t_i = 0.5(x_i + 1).$$

where $x_i'$'s are the Gaussian-Legendre Quadrature nodes in $(-1,1)$ and $a_i$'s are the corresponding weight. Substituting $\Delta(t_i)$ with $\hat{\Delta}(t_i)$, we estimate $\Delta_w$ as:

$$\hat{\Delta}_w = \sum_{i=1}^{m} q_i \hat{\Delta}(t_i),$$

$$t_1 = 0, t_m = 1, q_1 = -w(0), q_m = w(1), q_i = -0.5 a_i w'(t_i), 2 \leq i \leq m-1.$$

The variance of $\hat{\Delta}_w$ can be calculated in a similar way as for $\hat{\Delta}_{w,t}$. As an example,

$$Var[\hat{\Delta}_w] = \sum_{i=1}^{m} \sum_{j=1}^{m} q_i\, q_j\, \hat{\sigma}_{ij}.$$

Since $\Delta(0) = 0$ in an RCT, we can choose a smart $q_1^*$ similarly to how we estimate $\tilde{\Delta}_{w,t}$ for the discrete PPR:

$$q_1^* = -\sum_{i=2}^{m} q_i \frac{\hat{\sigma}_{i1}}{\hat{\sigma}_{11}},$$

which potentially leads to further precision improvement.

To apply this method, follow-up times need to follow the rule of Gaussian-Legendre Quadrature node. Under this rule, data collection schedule is not equally spaced. Instead, more frequent data collection is required at the beginning and end of the follow-up period than in the middle. A major advantage of the Gaussian-Legendre Quadrature method is that it provides some level of robustness (with sufficient number of nodes) to approximate the continuous PPR. As opposed to other parametric models, this method does not require strong assumption about $\Delta(t)$.

**2.4.3 Precision Improvement: Analytical Examples**

As a numerical example of precision improvement, we consider $\hat{\Delta}_{L,t}$ and $\hat{\Delta}_L$ for the situation where $Var(\hat{\Delta}(t)) = \sigma^2$ and $Cov\left(\hat{\Delta}(t), \hat{\Delta}(s)\right) = \tau$ $(t \neq s)$. In this case, it can be shown (see **Supplementary Material S2**) that

$$\frac{Var[\hat{\Delta}_{L,t}]}{Var[\hat{\Delta}_1]} = \frac{\sum_{i=1}^{m} v_i^2}{2}, \quad \frac{Var[\hat{\Delta}_L]}{Var[\hat{\Delta}_1]} = \frac{\sum_{i=1}^{m} q_i^2}{2}.$$

For $\hat{\Delta}_{L,t}$, under the special case of $m$ equally spaced time points, it can be shown that

$$\frac{Var[\hat{\Delta}_{L,t}]}{Var[\hat{\Delta}_1]} = \frac{6(m-1)}{m(m+1)},$$

which decreases with increasing $m$. In **Table 1**, we demonstrate variance reduction of $\hat{\Delta}_{L,t}$ and $\hat{\Delta}_L$ relative to $\hat{\Delta}_1$ across various values of $m$ (equal spaced time points for $\hat{\Delta}_{L,t}$). It is clear that the precision gain can be substantial as $m$ increases, yet still remains practically feasible. For instance, when $m = 8$, there are 42% and 15% reduction in the variance of $\hat{\Delta}_{L,t}$ and $\hat{\Delta}_L$ relative to $\hat{\Delta}_1$, respectively. If the treatment effect $\Delta_{L,t} = \Delta_L = \Delta_1$, these variance reductions translate to 42% and 15% reduction in sample size. For $\hat{\Delta}_{L,t}$, the precision under compound symmetric variance-covariance matrix can be further improved by placing time points closer to the beginning and end of the follow-up instead of spacing them equally. To see this, note that $\sum_{i=1}^{m} v_i^2 = 1/\sum_{i=1}^{m}(t_i - \bar{t})^2$. Thus, putting $t_i$'s closer to 0 or 1 can increase the denominator and subsequently reduce the variance of $\hat{\Delta}_{L,t}$.

When the variance or covariance of $\hat{\Delta}(t)$ is not constant, the variance reduction of $\hat{\Delta}_{L,t}$ and $\hat{\Delta}_L$ relative to $\hat{\Delta}_1$ depends on the specific variance-covariance structure. In certain scenarios, there can be an increase in the variance of $\hat{\Delta}_{L,t}$ and $\hat{\Delta}_L$. For instance, when $Var(\hat{\Delta}(t)) = \sigma^2$ and $Cov(\hat{\Delta}(t), \hat{\Delta}(s)) = \sigma^2 \times 0.5^{|t-s|}$, $Var[\hat{\Delta}_{L,t}]/Var[\hat{\Delta}_1] = 1.1$ for $m = 8$ equally spaced time points. In Section 3, we explored the variance reductions of alternative PPRs relative to $\hat{\Delta}_1$ under more scenarios.

## 2.5 Signal of Treatment Effect versus Estimation Precision

In Section 2.3, we proposed that considering a weight function $w(t)$ similar in shape to $\Delta'(t)$ could amplify the signal of treatment effect. In Section 2.4, we showed that certain choice of $w(t)$ can lead to improvement in estimation precision compared to using $w(t) \equiv 1$. This naturally raises the question of whether we can choose a $w(t)$ that optimizes the statistical power. However, we would like to emphasize that our goal is to create an interpretable summary measure of disease progression that can also improve statistical power, rather than to formulate the optimal test statistic. The optimal test, although mathematically possible, is fundamentally designed to be a test rather than an estimand. Consequently, it does not offer a meaningful summary measurement of disease progression or the difference between treatment groups. In addition, it requires knowledge of true values of $\Delta(t)$ and the variance-covariance matrix of its estimator (see Section 3 for more details). As such, it only bears theoretical value without much practical utility. Our position is that a pre-specified weight function, informed by the qualitative characteristics of the mean trajectories, can still substantially improve statistical power. To evaluate the trade-off/synergy between signal of treatment effect and estimation precision, we conducted numerical analyses in Section 3, studying sample size requirement for several weight functions discussed previously under several settings.

## 3. NUMERICAL EXAMPLES

In this section, we compared the performance of different discrete and continuous PPRs and their estimators. Our comparison focused on three aspects: 1) the signal of treatment effect; 2) the precision (standard error) of the PPR estimate; 3) sample size implications due to 1) and 2). All numerical examples were constructed under an RCT design.

**3.1 Design of the numerical settings**

*3.1.1 Trajectory of population means*

We fixed the population mean trajectory for the control arm and considered four distinct patterns of the treatment effect, $\Delta'(t)$, over a follow-up period of $0 < t \leq 1$ in an RCT setting. Specifically:

Mean trajectory for the control arm: $f(t) = -2.5t^3 + 6t^2 + 5t + 0.5$

Treatment effect: $\Delta'(t) =$

(i) $0.45(-1.1t + 2)^2$ (Monotonically decreasing)

(ii): $1$ (Constant)

(iii) $1.2(1 - e^{-6t})$ (Monotonically increasing)

(iv) $1.05\phi(t; 0.55, 0.25)$ (Increasing then decreasing).

Mean trajectory for the treatment arm: $h(t) = f(t) - \int_0^t \Delta'(x)dx$.

Here $\phi(t; 0.55, 0.25)$ denotes the normal probability density function with mean 0.55 and standard deviation 0.25. The trajectories of $f(t)$, $h(t)$, $\Delta(t)$ and their derivatives are shown in **Figure 2**. Note that for all four scenarios, $\Delta(1) = 1$.

*3.1.2 Variance-covariance matrix of outcome measure*

For discrete PPRs, we considered $m$ ($m$=5,6,...,10) equally spaced visits, where $t_1 = 0$ and $t_m = 1$. For a given $m$, the standard deviation of the outcome was defined as $\sigma_i = 1 + \frac{\sigma-1}{m-1}(i-1), i = 1,2,...,m$. This specification introduces an increasing trend in variation over time, aligned with common observations in practical situations. The correlation between repeated measures from the same individual was constructed with an autoregressive correlation matrix defined by: $\rho_{ij} = k\left(\frac{\rho}{k}\right)^{|t_i-t_j|}, \rho \leq k \leq 1, i \neq j$. This structure implies a decay in correlation over time, where measurements further apart show lower correlation. The parameter $k$ controls the variation of the correlations, e.g., smaller values of $k$ lead to more homogeneous correlations, while $k = \rho$ represents a compound symmetric correlation structure where all correlations are equal to $\rho$. Note that our setup ensures $\sigma_1 = 1$, $\sigma_m = \sigma$ and $\rho_{1m} = \rho$, which keeps the precision in estimating $\Delta(1) = 1$ constant regardless of $m$. Hence, the signal of treatment effect and precision in estimation using the mean CFB remain fixed in our experiment. In our numerical examples, we set $\rho = 0.6$ and considered $\sigma = (\sqrt{2}, \sqrt{3}, \sqrt{5})$ and $k = (0.6, 0.7, 0.8, 0.9)$. Specific examples of correlation matrix for different values of $k$ are provided in the **Supplementary Material S3**.

For continuous PPRs, we also considered $m$ ($m = 5, 6, \cdots, 10$) time points, with $m - 2$ visits corresponding to the Gaussian-Legendre Quadrature nodes. Therefore, the visits are not equally spaced. All parameter settings are identical to those specified for the discrete PPRs, except that $\sigma_i = 1 + (\sigma - 1)t_i$.

### 3.2 Choice of estimand and calculation of performance measures

We considered treatment effect as the absolute difference of three PPRs in our numerical experiments:

a) $\Delta_1$ (absolute difference in mean CFB),

b) $\Delta_{L,t}$ and $\Delta_L$ (absolute difference in discrete and continuous PPRs with the OLS weight function),

c) $\Delta_{A,t}$ and $\Delta_A$ (absolute difference in discrete and continuous PPRs with the AUC weight function).

To facilitate comparison of the signal of treatment effect, we used $\Delta_1 = 1$ as the reference and expressed the treatment effect measured by b) and c) as a percentage relative to $\Delta_1$. For precision comparison, we used the standard error of $\hat{\Delta}_1$ as the reference and expressed the standard error in estimating b) and c) as a percentage. The sample size requirement was compared based on the *signal-to-noise ratio* (SNR), defined as the square of the estimand of treatment effect divided by the variance of its estimator. For $\Delta_1$, the SNR was calculated as $\Delta_1^2/Var(\hat{\Delta}_1)$. Using $Var(\hat{\Delta}_1)/\Delta_1^2$ (i.e., inverse of the SNR for $\Delta_1$) as a reference, we expressed the corresponding quantity for b) and c) as a percentage. This percentage represented the relative sample size for b) and c) compared to $\Delta_1$.

For the comparison of sample size, we also included the optimal SNR, which corresponded to the smallest sample size for a test statistic that is linear in $\hat{\Delta}(t)$. Let $\Delta = (\Delta(t_1), \Delta(t_2), \ldots, \Delta(t_m))^T$, $\Sigma = (\sigma_{ij})$ $(i, j = 1, 2, \ldots, m)$ and $\hat{\Delta} = (\hat{\Delta}(t_1), \hat{\Delta}(t_2), \ldots, \hat{\Delta}(t_m))^T$, the SNR for the test statistic $T_{opt} = \Delta^T \Sigma^{-1} \hat{\Delta}$ is $\Delta^T \Sigma^{-1} \Delta$, providing the highest SNR among all test statistics that are linear in $\hat{\Delta}(t)$.[7] Therefore, $(\Delta^T \Sigma^{-1} \Delta)^{-1}$ expressed as a percentage of $Var(\hat{\Delta}_1)/\Delta_1^2$ represents the smallest sample size needed. Note that the optimal SNR is in general not achievable in practice, as the construction of this test statistic requires complete knowledge of $\Delta$ and $\Sigma$.

### 3.3 Numerical results

### 3.3.1 Estimands based on discrete PPRs

In **Figure 3**, the top panel compares the signal of the treatment effect among the three estimands across four progression trajectories as described in Section 3.1.1, while the bottom panel compares the standard error across values of $k = (0.6, 0.7, 0.8, 0.9)$ with $\sigma = \sqrt{2}$. Additional results for other values of $\sigma$ are provided in the **Supplementary Material S4**.

The results show no signal amplification when $\Delta'(t)$ is constant. Consistent with our derivation in Section 2.3, the signal of the treatment effect is amplified under scenarios where $w(t)$ and $\Delta'(t)$ are positively correlated. Specifically, $\Delta_{A,t}$ with a decreasing weight function yields approximately 20% signal amplification when $\Delta'(t)$ is decreasing, whereas $\Delta_{L,t}$ yields 8%-14% signal amplification when $\Delta'(t)$ follows an increasing-then-decreasing pattern. Conversely, the signal is reduced by 10% for $\Delta_{A,t}$ when $\Delta'(t)$ is increasing. Additionally, we observe that increasing the number of visits enhances the signal for $\Delta_{L,t}$ under increasing or increasing-then-decreasing $\Delta'(t)$, as well as for $\Delta_{A,t}$ when $\Delta'(t)$ is decreasing.

For the estimation of $\Delta_{L,t}$, the precision continues to improve as the number of visits increases for each fixed $k$ parameter. On the other hand, a smaller value of $k$ (i.e., less variation in the correlation coefficients between repeated measures) results in better precision for each fixed $m$. At the extreme situation with a compound symmetric correlation matrix ($k=0.6$), the precision improvement is the highest. However, unlike $\Delta_{L,t}$, estimation of $\Delta_{A,t}$ is less precise than $\Delta_1$ and the precision gets worse as $m$ increases.

**Figure 4** presents the relative sample size that is driven by both the signal of treatment effect and estimation precision. The minimal sample size based on the optimal test was included to provide a reference of the best possible scenario. Notably, $\Delta_{L,t}$ requires the smallest sample sizes across almost all scenarios, except when $\Delta'(t)$ is a decreasing function and the variation of $\rho_{ij}$ is large (i.e., $k=0.8, 0.9$). In particular, under increasing or increasing-then-decreasing $\Delta'(t)$, the required sample size for $\Delta_{L,t}$ is nearly identical to the minimal sample size. Under these two trajectory patterns, $\Delta_{L,t}$ achieves approximately 50% and 60% sample size reduction relative to $\Delta_1$ for $k=0.6$ and $m=10$, respectively. For $\Delta_{L,t}$, sample size consistently decreases with more study visits. In contrast, $\Delta_{A,t}$ generally requires more sample size than $\Delta_1$, except when $\Delta'(t)$ is a decreasing function. The saving in sample size under this trajectory setting, as discussed earlier, is exclusively driven by signal amplification. $\Delta_{A,t}$ has an advantage over $\Delta_{L,t}$ only when $\Delta'(t)$ is a decreasing function and the variation of $\rho_{ij}$ is large (i.e., $k=0.8, 0.9$).

### 3.3.2 Estimands based on continuous PPRs

The results for estimands based on continuous PPRs and Gaussian-Legendre Quadrature estimation procedure (**Figures 5 and 6**) are similar to those for the discrete PPRs with a few exceptions. First, unlike the discrete PPRs, signal amplification for $\Delta_L$ and $\Delta_A$ is fairly stable with respect to $m$. This is expected as $\Delta_L$ and $\Delta_A$ does not depends on $m$ and the Gaussian-Legendre should work quite well as *m* gets reasonably large. On the other hand, $\Delta_{L,t}$ and $\Delta_{A,t}$ based on the discrete PPRs do depend on *m*. Second, the precision improves with increasing number of visits for both $\Delta_L$ and $\Delta_A$. However, $\Delta_L$ outperforms $\Delta_1$ only when the number of visits is relatively large and when the variation of $\rho_{ij}$ is small (e.g., *m*>7 for *k*=0.6,0.7 or *m*>8 for *k*=0.8). Third, the advantages of $\Delta_L$ over $\Delta_1$ in sample size requirement remains clear for all trajectories except when $\Delta'(t)$ is a decreasing function. But it achieves the advantage with only higher value of *m*. Lastly, $\Delta_A$ also needs higher value of *m* and large variation in $\rho_{ij}$ (i.e., *k*=0.8,0.9) to achieve sample size advantage over $\Delta_1$ when $\Delta'(t)$ is a decreasing function.

### 3.3.3 Estimation using $v_1^*$ and $q_1^*$

We also evaluated estimation using the "smart" coefficients $v_1^*$ and $q_1^*$ as described in Sections 2.4.1 and 2.4.2. (see **Supplementary Materials S5 and S6**). As expected, the standard error was reduced compared to the results from Sections 3.3.1 and 3.3.2. The reduction was more pronounced for the estimation of $\Delta_{A,t}$ than the other discrete PPR estimands. The same observation also holds for continuous PPRs estimands.

## 4. APPLICATION TO A REAL DATA

The EMERGE trial is a randomized, double-blinded, placebo-controlled, global, phase 3 study of aducanumab in patients with early Alzheimer's disease (AD) [NCT02484547].[8] It included 1638 subjects with confirmed amyloid beta pathology, aged 50 to 85 years, who met clinical criteria for mild cognitive impairment due to AD or mild AD dementia. The study comprised a placebo-controlled period of 78 weeks with subjects receiving monthly low dose of aducanumab (3 or 6 mg/kg target dose), high dose (10mg/kg target dose), or placebo followed by a dose-blind long-term extension (LTE) period with subjects receiving monthly dose of aducanumab until the end of the study, lost to follow-up or early trial termination (March 21, 2019), whichever occurred first. The primary COA was Clinical Dementia Rating-Sum of Boxes (CDR-SB), which was measured at baseline, week 26, week 50, and week 78 during the placebo-controlled period, and was scheduled to be measured at week 106, week 134, week 162, week 182, week 232, week 284, and week 338 during the LTE period.

We focused on estimating treatment effect of high dose aducanumab as measured by $\Delta_1$, $\Delta_{L,t}$ and $\Delta_{A,t}$ during two follow-up periods: 1) placebo-controlled period (78 weeks of follow-up) with *m*=4; 2) placebo-controlled period plus first clinical visit during LTE (106 weeks of follow-up) with *m*=5. Because subjects initially randomized to the placebo arm started receiving either low or high aducanumab between week 78 and 106, the treatment effect in 2) reflects the benefit of starting aducanumab early.

MMRM was used to estimate the mean CDR-SB at each visit for the high dose and placebo groups, using all available data collected up to visit 78/106, including treatment group, visit, treatment group by visit interaction, baseline Mini Mental State Examination (MMSE), AD symptomatic medication used at baseline, region (United States, Europe/Canada/Australia), and ApoE $\varepsilon$4 status (carrier/non-carrier) as covariates. An unstructured covariance matrix was used to account for the correlation across the repeated measurements. The difference of the mean CDR-SB between the placebo and high dose group at each visit, as well as its variance-covariance matrix was estimated from the MMRM.

MMRM estimates of CDR-SB trajectories for the placebo and high-dose aducanumab groups were shown in **Figure 7.** We can see that the difference between the placebo and high dose aducanumab group was small at week 26, during which a titration regimen (1mg/kg for the first 2 doses, 3mg/kg for the next 2 doses, 6 mg/kg for the next 2 doses, and 10mg/kg thereafter) was implemented. Following that, the treatment started to slow progression until week 78, with a clear increase in slowing progression speed between week 50 and 78 as compared with between week 26 and 50. After week 78, subjects initially randomized to the placebo arm started receiving either low or high aducanumab (i.e., delayed treatment). As expected, the treatment effect then attenuated from week 78 to week 106 during the LTE period with similar progression rates between the two arms.

Estimates of the three PPR estimands of treatment effect were summarized in **Figure 8** with $\hat{\Delta}_1$ as the reference. For treatment effect during the 78 weeks of placebo-controlled period, $\hat{\Delta}_{L,t}$ had a similar point estimate and standard error as $\hat{\Delta}_1$ (96% and 99%, respectively), leading to similar SNR: $Z^2 = \left(\frac{Point\ estimate}{Standard\ error}\right)^2$. In contrast, $\hat{\Delta}_{A,t}$ had a reduced treatment effect, with lower signal and precision, leading to a 66% reduction in $Z^2$ compared to $\hat{\Delta}_1$. For treatment effect from baseline to week 106, $\hat{\Delta}_{L,t}$ showed a notable improvement with a 76% increase in $Z^2$ compared with $\hat{\Delta}_1$, largely driven by a 28% increase in point estimate (signal of treatment effect). $\hat{\Delta}_{A,t}$ had a similar $Z^2$ to $\hat{\Delta}_1$, due to a 10% increase in signal but an 8% increase in standard error.

## 5. DISCUSSION

In this article, we proposed the PPR as a general framework for summarizing disease progression during the follow-up period of a clinical study. As an estimand, the PPR is a weighted average of the local or instantaneously slope of the trajectory of the population mean of the outcome. The flexible choice of the weight allows us to emphasize or deemphasize certain period of follow-up time. It potentially enhances the signal-to-noise ratio when detecting the treatment effect, which could substantially improve statistical power. The PPR is a generic concept that includes a variety of intuitive options such as change from baseline, area under the curve, and the slope of an ordinary least-square fit to the trajectory. We focus our numerical studies on estimating treatment effect based on these three types of PPRs due to their natural interpretability and potential to enhance the signal of treatment effect and precision.

One practical advantage of the PPR framework is that statistical inference is quite straightforward. For instance, parameter estimates and corresponding variance-covariance matrix from MMRM outputs from standard statistical software are sufficient to compute point estimate and the standard error of a PPR or the difference of PPRs between two comparison groups. It should be noted that directly fitting a linear slope to the data through a longitudinal model (e.g., linear mixed-effects model) in general do not yield valid inference for $r_{L,t}$ or $\Delta_{L,t}$. This is because missing-data and correlation structure assumed in these models could introduce bias in the slope estimator.

For a specific RCT, the choice of the weight function should ideally strike a balance between signal of treatment effect and precision as discussed in Section 2.5. Our numerical studies suggest that $r_{L,t}/r_L$ offers a good performance when there is a delay in the manifestation of the treatment effect, while $r_{A,t}/r_A$ offers some benefit when there is an immediate treatment effect. Importantly, we also need to factor in two considerations when selecting the weight function. First, the weight function should lead to an interpretable PPR. In this regard, both $r_{L,t}/r_L$ and $r_{A,t}/r_A$ offer appealing interpretations. Second, to ensure that treatment effects are comparable across multiple RCTs, the same weight function needs to be used across these trials. Discrete PPRs may lead to incomparability if different numbers or locations of time points are used in different trials. The continuous PPR is more robust, but it still requires consistent use of the same weight function in each trial.

Although we focus on the PPRs with "ordinary least-square" and "area under the curve" weight functions, there are other weight functions to be considered in practice. Some examples of the weight function for continuous PPR include:

a) Members of the class of Beta distributions

b) $w(t) \propto min\{1 - t, 0.5\} \to r_w \propto \int_{0.5}^{1}[f(t) - f(0)]dt$: partial area under the curve

c) $w(t) \propto 1 - t^\alpha \to r_w \propto \alpha \int_{0}^{1} t^{\alpha-1}[f(t) - f(0)]dt$: weighted area under the curve.

There is a special feature of the class of Beta distributions in a). If $w(t) = Beta(t; a, b)$ ($a \geq 1, b \geq 1$), then $r_w$ is the slope of a weighted least-square linear fit to $f(t)$ with the weight function being $Beta(t; a - 1, b - 1)$ (see **Supplementary Material S7**). A new weight function can also be created by taking the weighted average of two weight functions. For example, $w(t) = 0.5 + 0.5 \times 6t(1 - t)$ is the average (with equal weight) of the weight of $r_1$ and $r_L$. This weight function has less variation among the weights assigned to different time points.

For the calculation of $\hat{\Delta}_w$ based on Gaussian-Legendre Quadrature approach, $t_2$ and $t_{m-1}$ could be fairly close to baseline $t_1 = 0$ and end of follow-up $t_m = 1$, respectively. For instance, when $m = 8$, $t_2 - t_1 = t_8 - t_7 = 0.034$. There could be a practical difficulty of requiring study participants to have their COAs collected twice in short time intervals at the beginning and end of a study, particularly for costly data and/or data that requires a clinical visit. On the other hand, because these time intervals between the two data collections are quite short, it may be reasonable to assume that $\Delta(t_2) - \Delta(t_1)$ and $\Delta(t_m) - \Delta(t_{m-1})$ are close to zero. Therefore, we can design a study with $m = 8$ but do not actually collect data at $t_2$ and $t_7$ and set $\hat{\Delta}(t_2) = \hat{\Delta}(t_1)$ and $\hat{\Delta}(t_7) = \hat{\Delta}(t_8)$ in the analysis.

An alternative measure of treatment effect based on the PPR is the ratio of the PPR under treatment to the PPR under control: $R_w = r_w(h)/r_w(f)$. Wang et al. considered the "proportional effect of the treatment" in AD, where the progression speed under treatment is assumed to be a constant proportion of the progression speed under control.[9] They termed this proportion Cognitive Progression Ratio (CPR). In our notation, the assumption translates to $h'(t) = CPR \times f'(t)$ ($0 \le t \le 1$), which implies that $R_w \equiv CPR$ does not depend on the weight function.

While the definition of PPR in this article is motivated by characterization of the longitudinal trajectory of the population mean of a continuous outcome, the PPR can in fact be defined for any monotone curve that summarizing disease progression for a population. For instance, the trajectory could represent the population median instead of population mean. Notably, there is a connection between PPR and similar concepts in time-to-event analysis. First, if $f$ represents the cumulative distribution function of the event time, then $r_1(f)$ and $r_A(f)$ correspond to the event rate at the end of the follow-up and twice the Restricted Mean Loss Time (RMLT),[10] respectively. Second, if $f(t)$ and $h(t)$ denote the negative of the logarithm of the survival function of the event time for control and treatment arms, respectively, then $\Delta_w$ becomes the weighted average of the difference in hazard function. This is in principle similar to the rationale behind the weighted log-rank test.[11] Furthermore, under the proportional hazard assumption, the hazard ratio is equal to $r_w(h)/r_w(f)$ for any arbitrary weight function $w(t)$.

Although we focus on the concept and estimation of PPR in the context of panel data from an RCT, they can also be extended to longitudinal studies without randomization or/and those with unequal follow-up time points. For observational longitudinal studies, established causal inference approaches can be used to adjust for potential confounding and estimate the potential outcome trajectories, which can then be used for the inference of PPR. In studies with varying follow-up time points among participants, the continuous version of PPR is more appropriate. Its estimation may require assumptions about $f(t)$ to achieve a good balance between precision and bias. An important feature of the PPR is that certain members, such as $r_{L,t}$ or $r_L$, could benefit from more time points in terms of enhancing the signal of treatment effect and/or precision in estimation, as demonstrated in the numerical studies in Section 3. Recent development of digital biomarkers and endpoints that can be collected at a much higher frequency than conventional COAs offers great potential to further improve the accuracy of treatment effect estimates using PPR.

## Acknowledgements

We would like to thank Dr. Fabian Buller from Neurimmune for insightful suggestions to improve the clarity of the presentations in this manuscript.

## Conflict of Interest

**Table 1** Ratio of variances of $\hat{\Delta}_{L,t}$ (equal space) and $\hat{\Delta}_L$ with respect to variance of $\hat{\Delta}_1$

| m | $\dfrac{\text{Var}[\hat{\Delta}_{L,t}]}{\text{Var}[\hat{\Delta}_1]}$ | $\dfrac{\text{Var}[\hat{\Delta}_L]}{\text{Var}[\hat{\Delta}_1]}$ |
|---|---|---|
| 5 | 0.80 | 1.67 |
| 6 | 0.71 | 1.25 |
| 7 | 0.64 | 1.01 |
| 8 | 0.58 | 0.85 |
| 9 | 0.53 | 0.74 |

**Figure Legend**

**Figure 1** Weight functions for various PPRs

**Figure 2** Left panels: trajectory of the population means of the outcome under the control arm (red curve) and the treatment arm (blue curve), and the difference in the population mean (grey curve) between the control and treatment arms. Right panels: Instantaneous slopes of the trajectories and their difference.

**Figure 3** Relative difference in signal of treatment effect (top panels) and standard error (bottom panels) for discrete PPRs. The mean CFB serves as the reference and is always 100%.

**Figure 4** Relative sample size for discrete PPRs. The mean CFB serves as the reference and is always 100%.

**Figure 5** Relative difference in signal of treatment effect (top panels) and standard error (bottom panels) for continuous PPRs. The mean CFB serves as the reference.

**Figure 6** Relative sample size for continuous PPRs. The mean CFB serves as the reference.

**Figure 7** Adjusted mean CDR-SB score by treatment group using MMRM for longitudinal data measured from baseline to Week 106. Subjects in the high dose group received high dose aducanumab from baseline to Week 106 (solid purple line), whereas subjects in the placebo group remained untreated from baseline to Week 78 (solid grey line) with delayed aducanumab treatment between Week 78 to 106 (dashed blue line).

**Figure 8** Comparison in (a) signal of treatment effect, (b) standard error and (c) signal to noise ratio for different discrete PPRs using the EMERGE clinical trial with data measured from baseline to Week 78 and to Week 106.

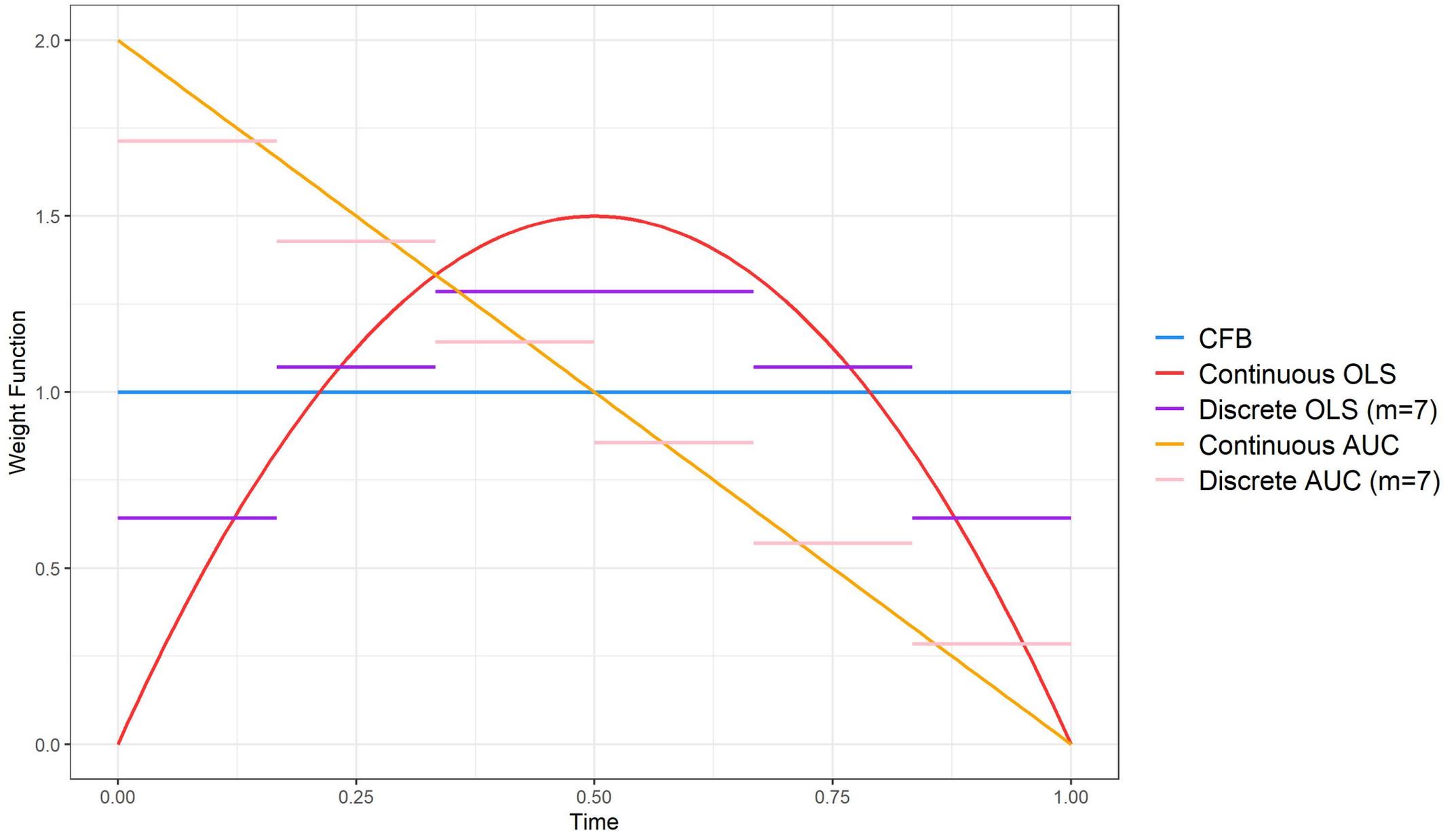

Changyu Shen, Figure 1

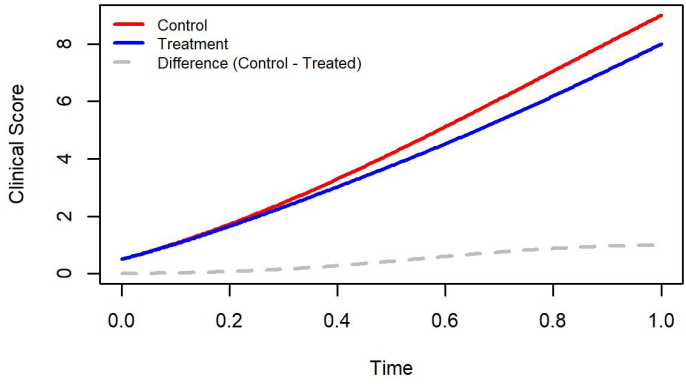
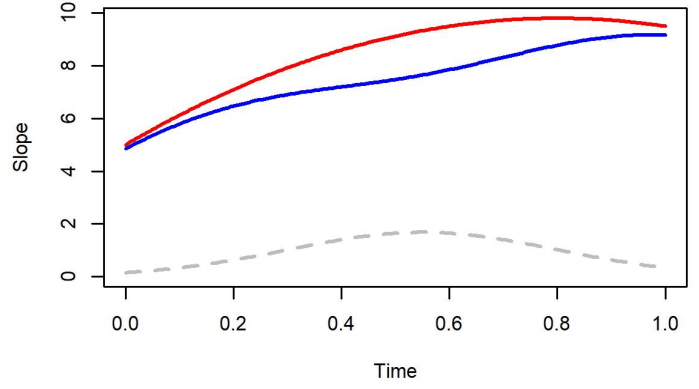

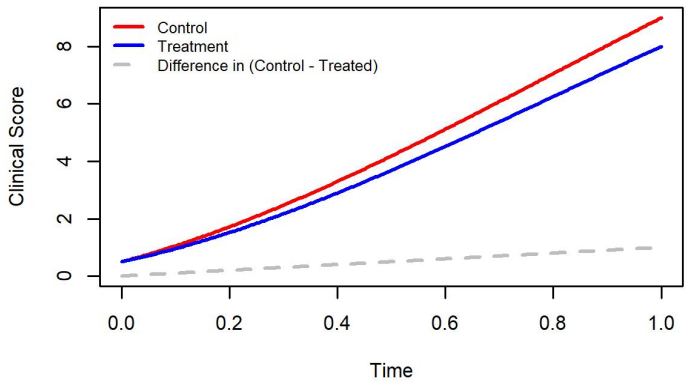
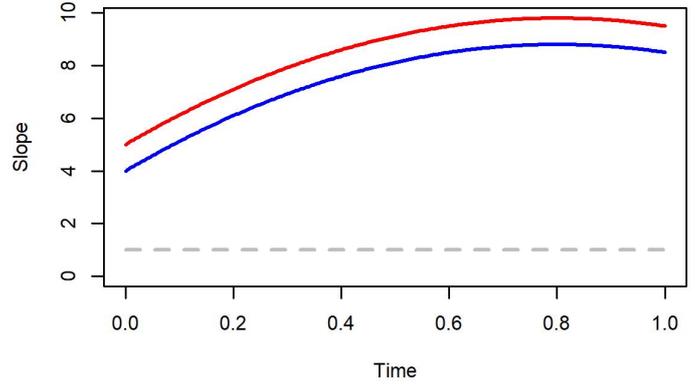

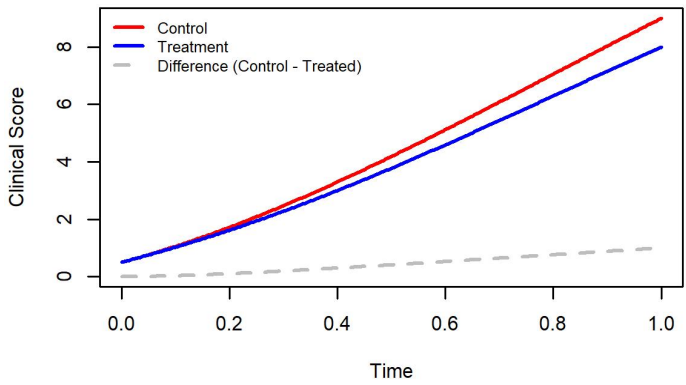
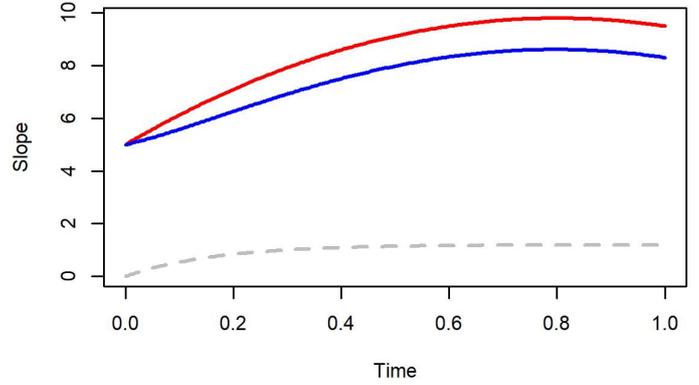

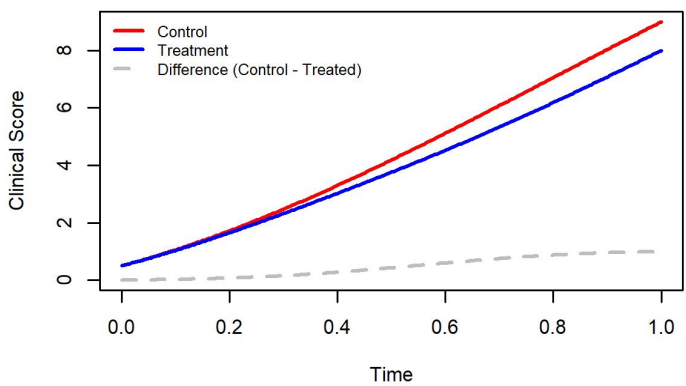
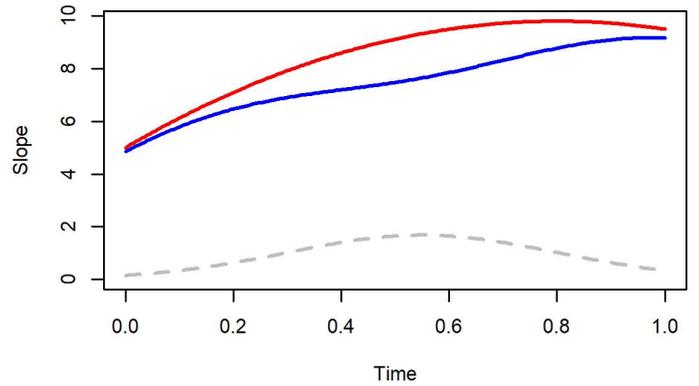

Changyu Shen, Figure 2

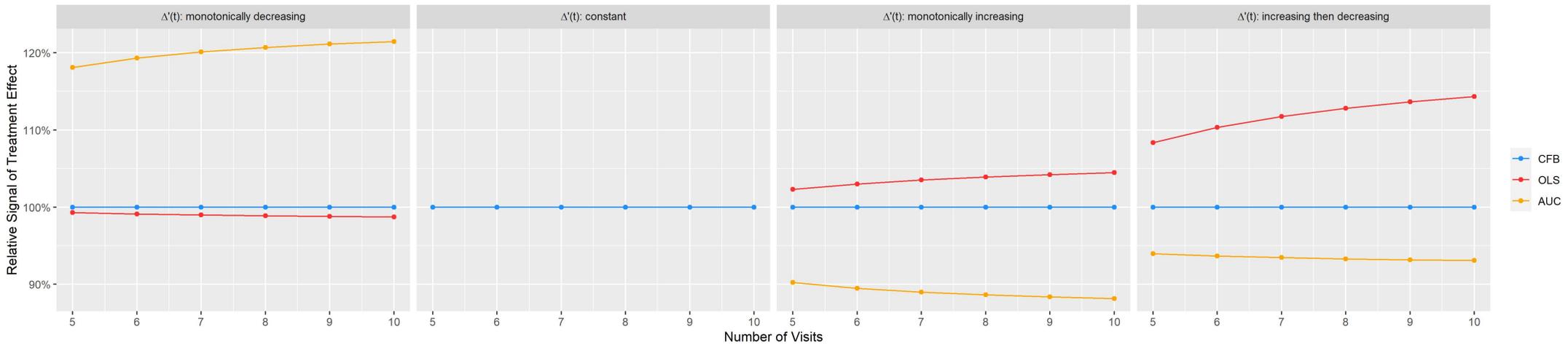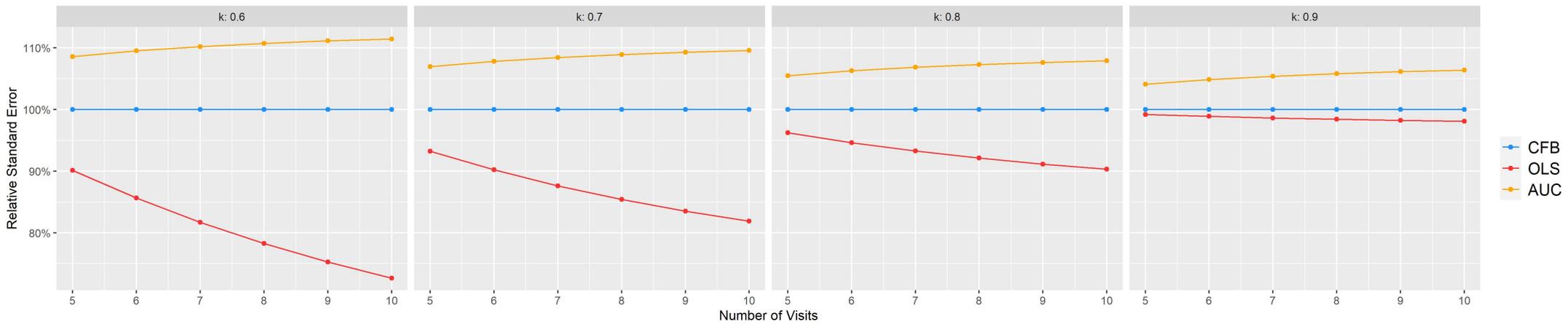

Changyu Shen, Figure 3

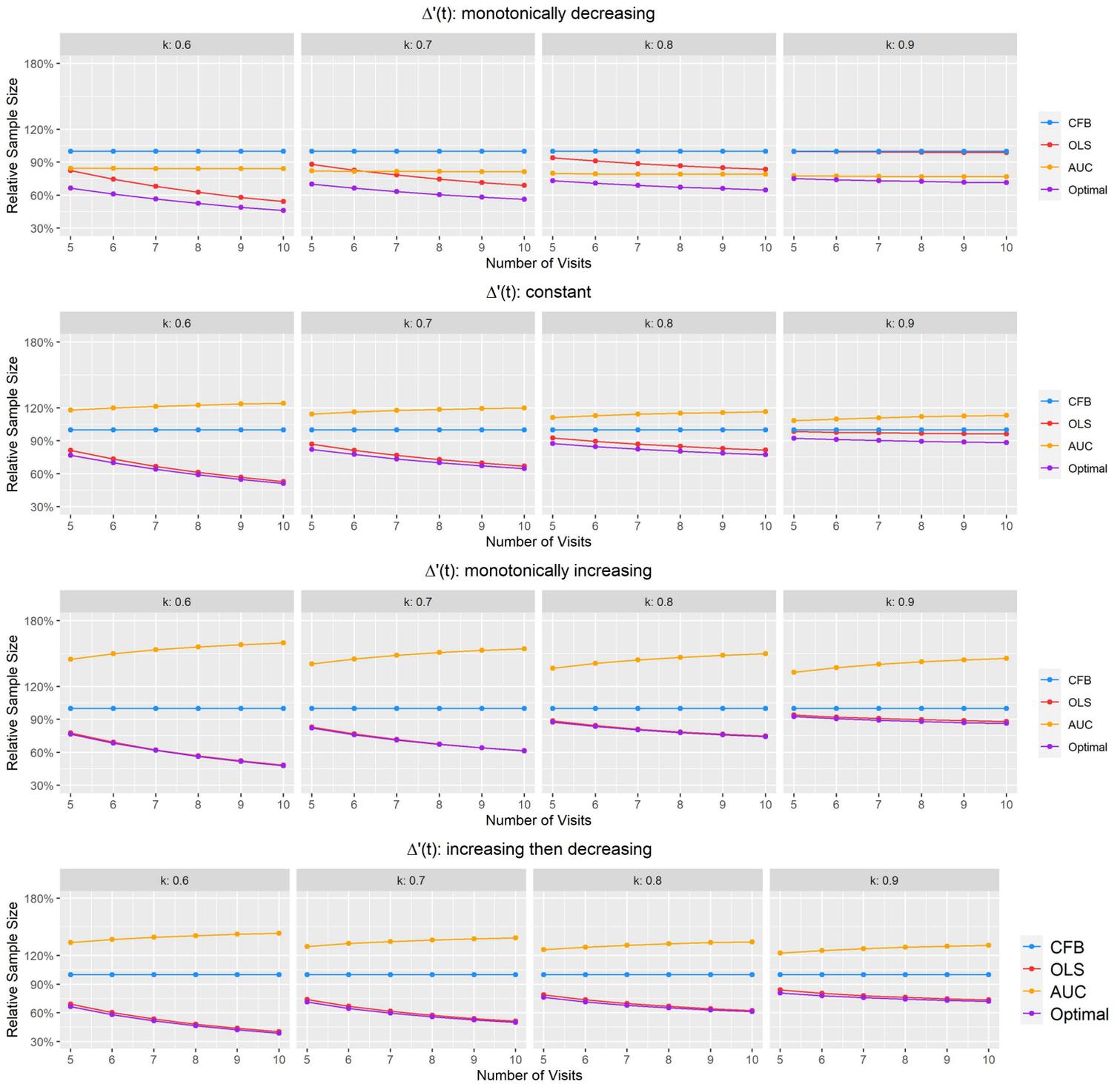
Changyu Shen, Figure 4

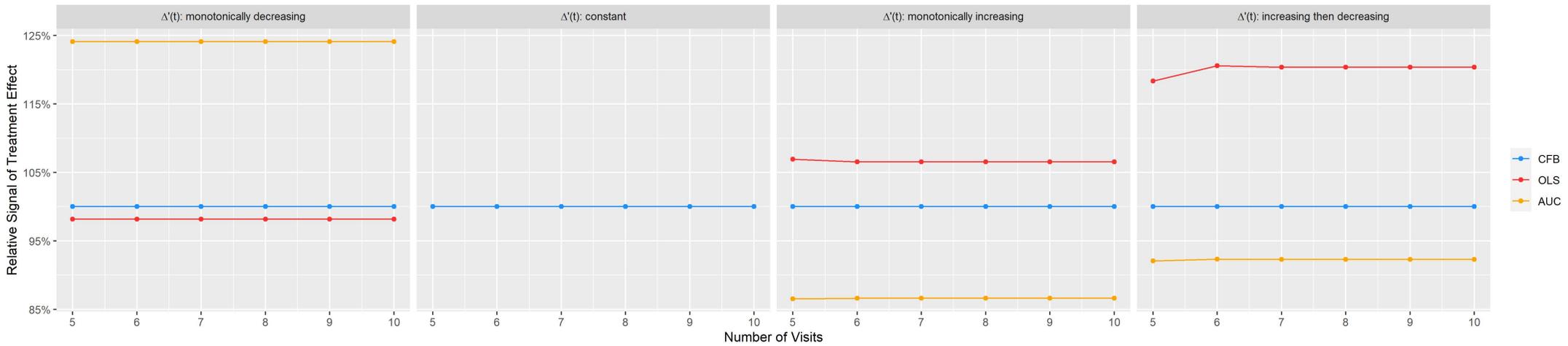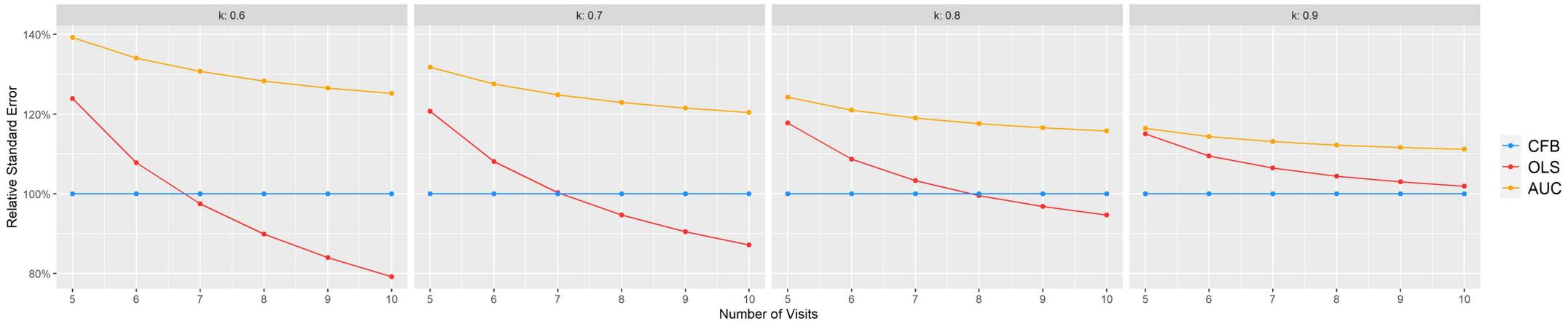

Changyu Shen, Figure 5

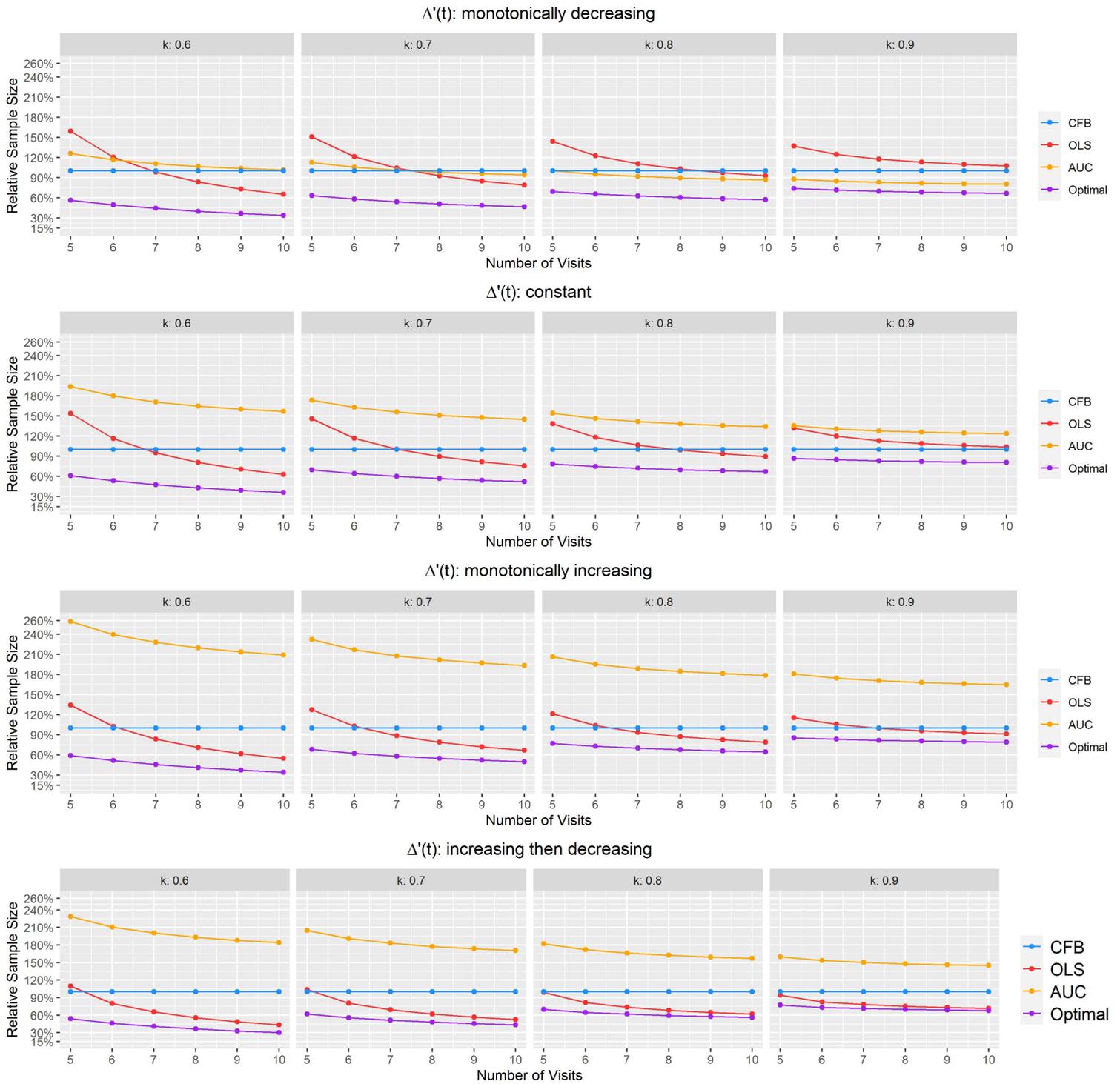

Changyu Shen, Figure 6

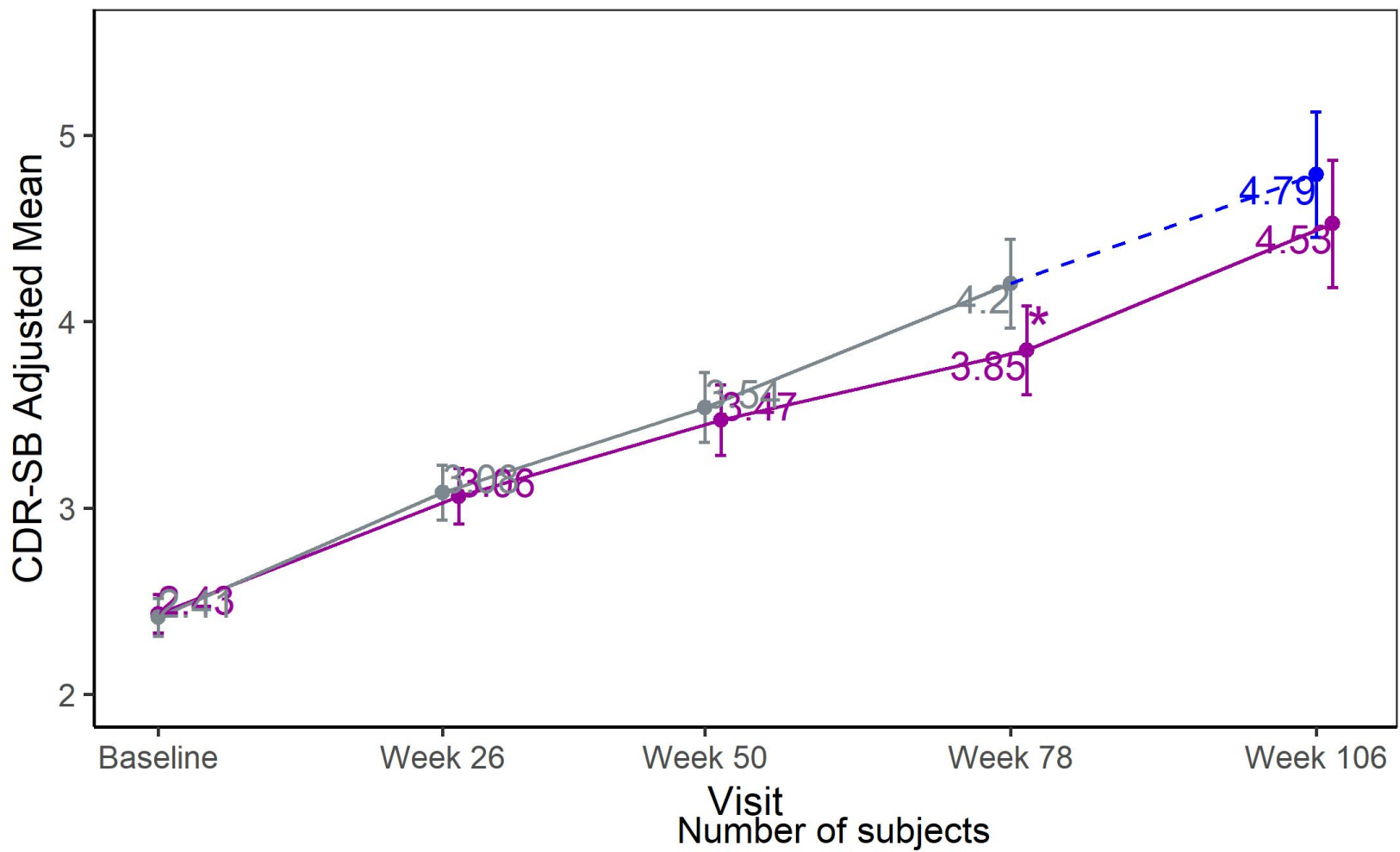

Changyu Shen, Figure 7

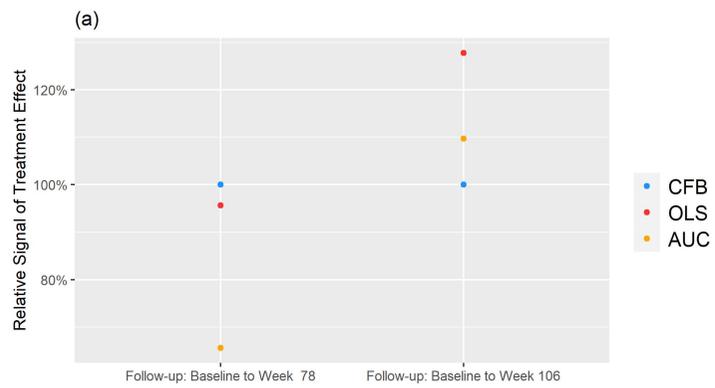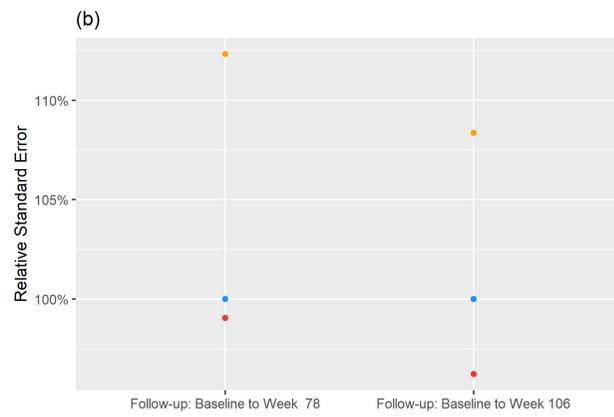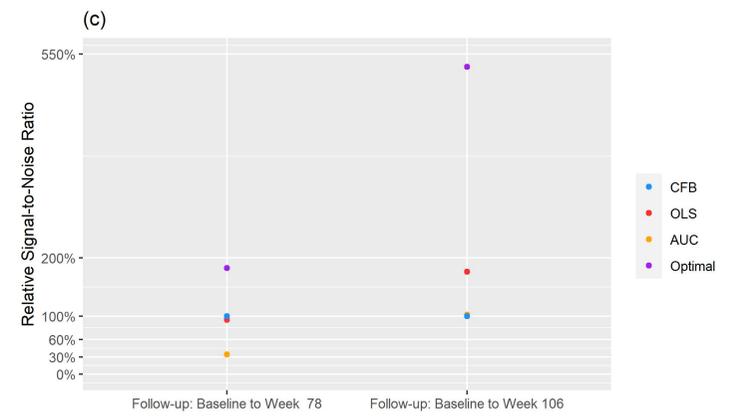

Changyu Shen, Figure 8

# Supplementary Materials for Manuscript: Using Principal Progression Rate to Quantify and Compare Disease Progression in Randomized Controlled Trials

Author: Changyu Shen, Menglan Pang, Ling Zhu, Lu Tian

## S1. Derivation of $v_1^*$

Let $\hat{\theta} = \sum_{i=2}^{m} v_i \hat{\Delta}(t_i)$, $\varepsilon^2 = Var(\hat{\theta})$ and $\gamma = Cov\left(\hat{\theta}, \hat{\Delta}(t_1)\right) = \sum_{i=2}^{m} v_i \sigma_{i1}$. Let $\hat{\Delta}(a) = \hat{\theta} + a\hat{\Delta}(t_1)$ be a class of estimators for $\Delta_{w,t}$. Then

$$Var\left(\hat{\Delta}(a)\right) = \sigma_{11} a^2 + 2\gamma a + \varepsilon^2 = \sigma_{11}\left(a + \frac{\gamma}{\sigma_{11}}\right)^2 + \varepsilon^2 - \frac{\gamma^2}{\sigma_{11}}.$$

Thus, $Var\left(\hat{\Delta}(a)\right)$ reaches minimum at $a^* = -\frac{\gamma}{\sigma_{11}}$. Replacing $\sigma_{i1}$ and $\sigma_{11}$ with $\hat{\sigma}_{i1}$ and $\hat{\sigma}_{11}$, we have

$$v_1^* = -\sum_{i=2}^{m} v_i \frac{\hat{\sigma}_{i1}}{\hat{\sigma}_{11}}.$$

## S2. Relative efficiency of $\hat{\Delta}_{L,t}$ and $\hat{\Delta}_L$ to $\hat{\Delta}_1$ under compound symmetric variance-covariance matrix

When $\sigma_{ii} = \sigma$ and $\sigma_{ij} = \tau$, we have

$$\widehat{Var}[\hat{\Delta}_{L,t}] = \sigma \sum_{i=1}^{m} v_i^2 + \tau \sum_{i \neq j} v_i v_j.$$

Because $\sum_{i=1}^{m} v_i = 0 \to \left(\sum_{i=1}^{m} v_i\right)^2 = 0$, we have $\sum_{i \neq j} v_i v_j = -\sum_{i=1}^{m} v_i^2$. Therefore,

$$\widehat{Var}[\hat{\Delta}_{L,t}] = (\sigma - \tau) \sum_{i=1}^{m} v_i^2.$$

For Gaussian-Legendre Quadrature approximation of $\Delta_L$, note that $q_1 = q_m = 0$. The nodes of Gaussian-Legendre Quadrature $x_i'$s are symmetric around 0. As a result, $t_i = 0.5(x_i + 1)$ is symmetric around 0.5. Since $L'(t) = 6(1 - 2t)$, $L'(t_i) = -L'(t_{m-i+1})$ ($2 \leq i \leq m - 1$). Because $a_i = a_{m-i+1}$ ($2 \leq i \leq m - 1$) for Gaussian-Legendre Quadrature, $\sum_{i=2}^{m-1} q_i = 0$. Therefore, $\sum_{i=1}^{m} q_i = 0$. As a result,

$$\widehat{Var}[\hat{\Delta}_L] = (\sigma - \tau) \sum_{i=1}^{m} q_i^2.$$

Finally, $\hat{\Delta}_1 = \hat{\Delta}(t_m) - \hat{\Delta}(t_1)$

$$Var(\hat{\Delta}_1) = 2(\sigma - \tau).$$

## S3. Example of correlation matrix for different values of $k$

An autoregressive correlation matrix between two repeated measures from the same individual was specified for the numerical examples: $\rho_{ij} = k \left(\frac{\rho}{k}\right)^{|t_i - t_j|}, \rho \leq k, i \neq j, i,j = 1,2,\ldots,m$, and $\rho = 0.6$. The value of $k$ controls the variation of the correlations, e.g. smaller value of $k$ leads to more homogeneous correlations. All correlations are equal to $\rho$ in the extreme case of $k=\rho$, representing a compounding symmetric correlation structure.

For $k = 0.6$, the correlation matrix is:

$$\begin{bmatrix} 1 & 0.6 & 0.6 & 0.6 & 0.6 \\ 0.6 & 1 & 0.6 & 0.6 & 0.6 \\ 0.6 & 0.6 & 1 & 0.6 & 0.6 \\ 0.6 & 0.6 & 0.6 & 1 & 0.6 \\ 0.6 & 0.6 & 0.6 & 0.6 & 1 \end{bmatrix}$$

For $k = 0.7$, the correlation matrix is:

$$\begin{bmatrix} 1 & 0.67 & 0.65 & 0.62 & 0.6 \\ 0.67 & 1 & 0.67 & 0.65 & 0.62 \\ 0.65 & 0.67 & 1 & 0.67 & 0.65 \\ 0.62 & 0.65 & 0.67 & 1 & 0.62 \\ 0.6 & 0.62 & 0.65 & 0.67 & 1 \end{bmatrix}$$

For $k = 0.8$, the correlation matrix is:

$$\begin{bmatrix} 1 & 0.74 & 0.69 & 0.64 & 0.6 \\ 0.74 & 1 & 0.74 & 0.69 & 0.64 \\ 0.69 & 0.74 & 1 & 0.74 & 0.69 \\ 0.64 & 0.69 & 0.74 & 1 & 0.74 \\ 0.6 & 0.64 & 0.69 & 0.74 & 1 \end{bmatrix}$$

For $k = 0.9$, the correlation matrix is:

$$\begin{bmatrix} 1 & 0.81 & 0.73 & 0.66 & 0.6 \\ 0.81 & 1 & 0.81 & 0.73 & 0.66 \\ 0.73 & 0.81 & 1 & 0.81 & 0.73 \\ 0.66 & 0.73 & 0.81 & 1 & 0.81 \\ 0.6 & 0.66 & 0.73 & 0.81 & 1 \end{bmatrix}$$

## S4 Additional numerical results for the estimates based on discrete PPRs

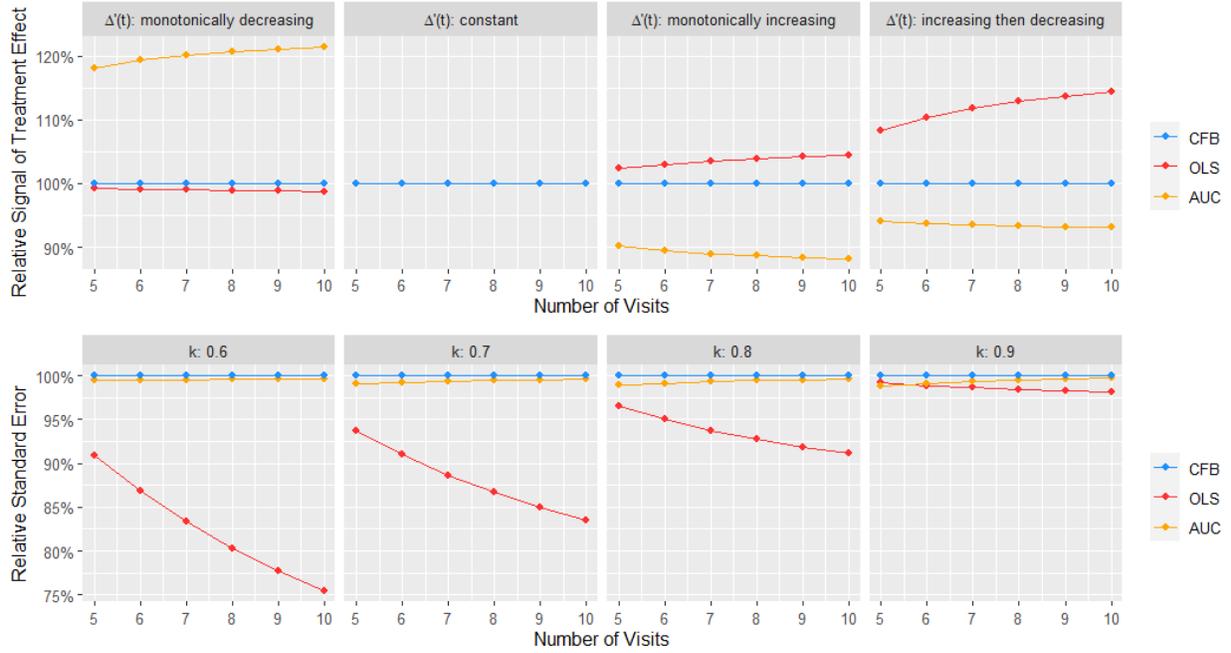

**Figure S1** Relative difference in signal of treatment effect (top panels) and standard error (bottom panels) for discrete PPRs. The CFB serves as the reference. Results based on $\sigma = \sqrt{3}$.

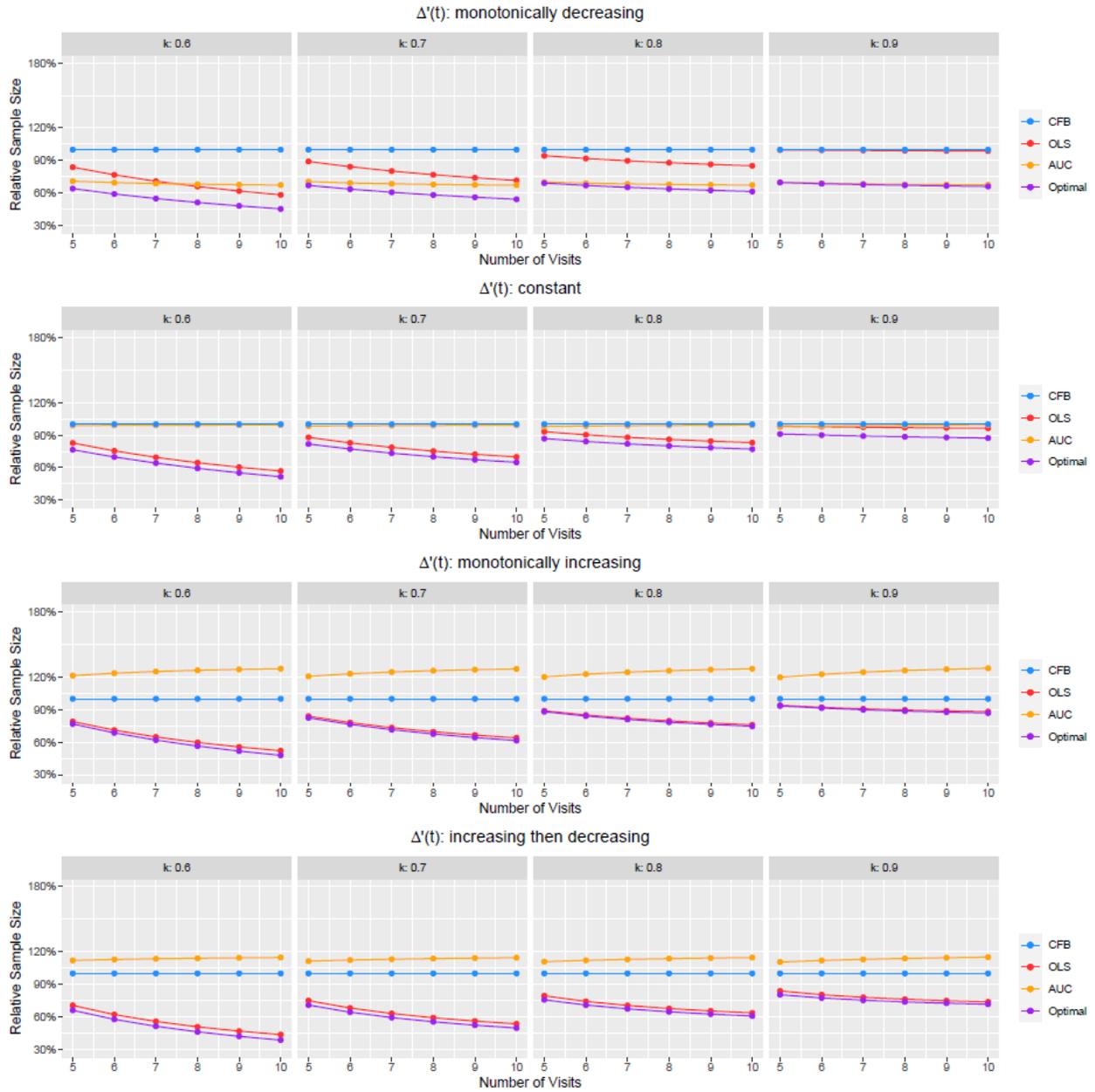

**Figure S2** Relative sample size for discrete PPRs. The CFB serves as the reference. Results based on $\sigma = \sqrt{3}$.

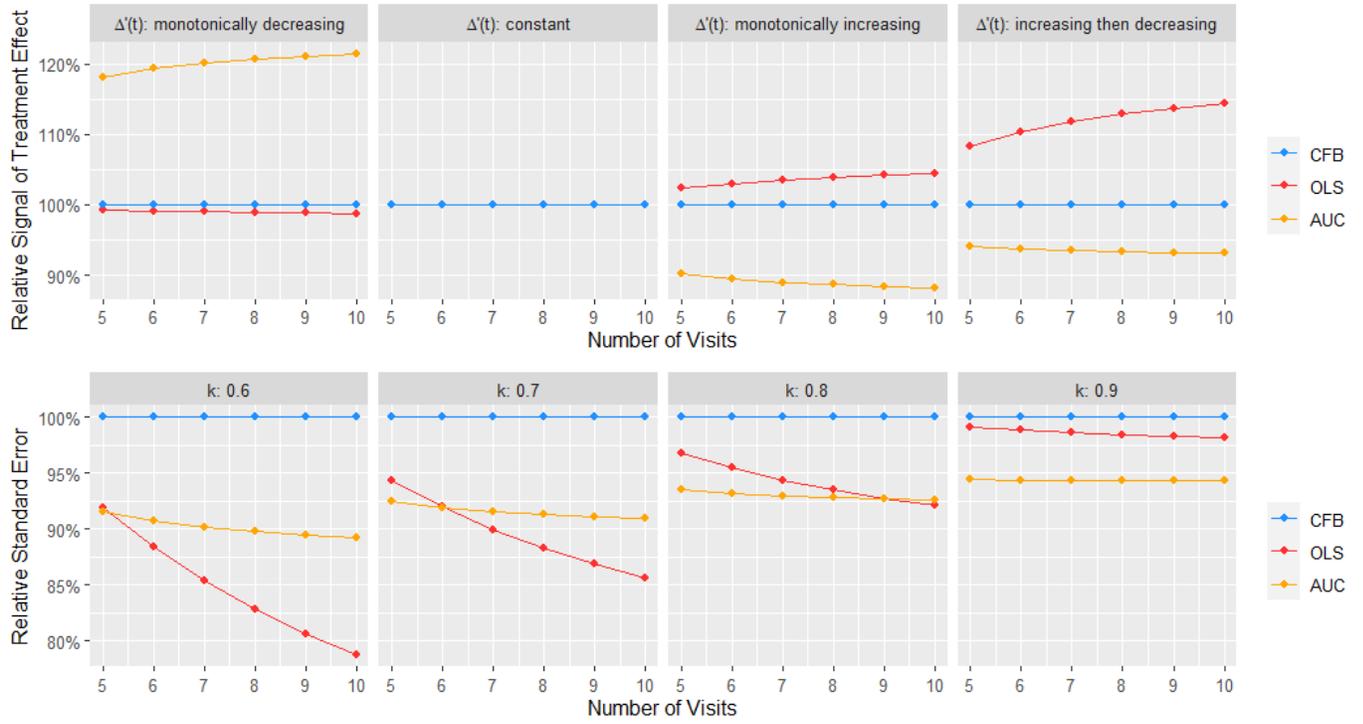

**Figure S3** Relative difference in signal of treatment effect (top panels) and standard error (bottom panels) for discrete PPRs. The CFB serves as the reference. Results based on $\sigma = \sqrt{5}$.

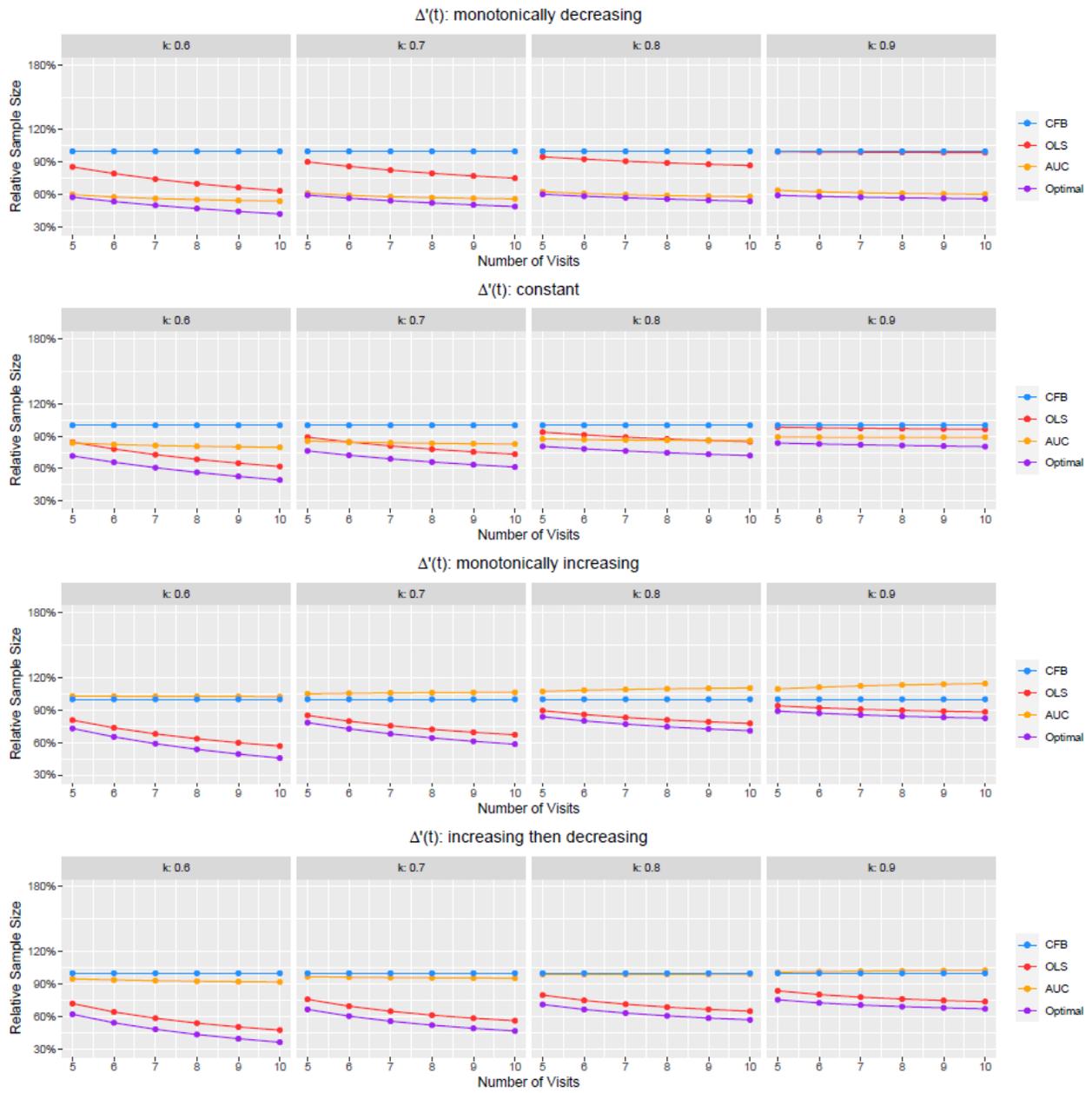

**Figure S4** Relative sample size for discrete PPRs. The CFB serves as the reference. Results based on $\sigma = \sqrt{5}$.

## S5. Numerical results for the estimates based on discrete PPRs using the "smart" coefficients $v_1^*$

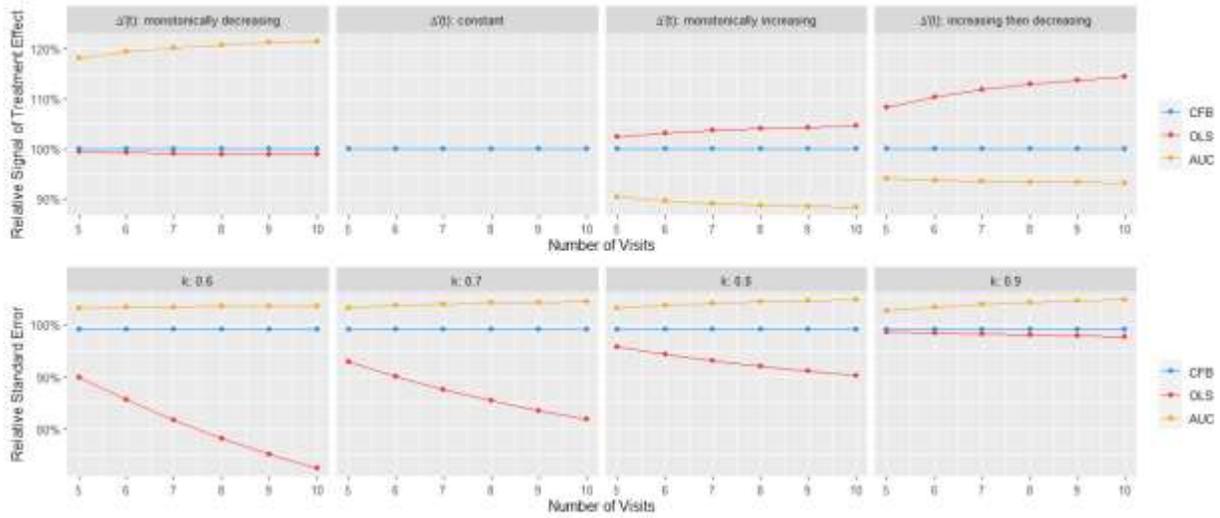

**Figure S5** Relative difference in signal of treatment effect (top panels) and standard error (bottom panels) for discrete PPRs. The CFB serves as the reference. Results based on "smart" coefficients $v_1^*$ and $\sigma = \sqrt{2}$.

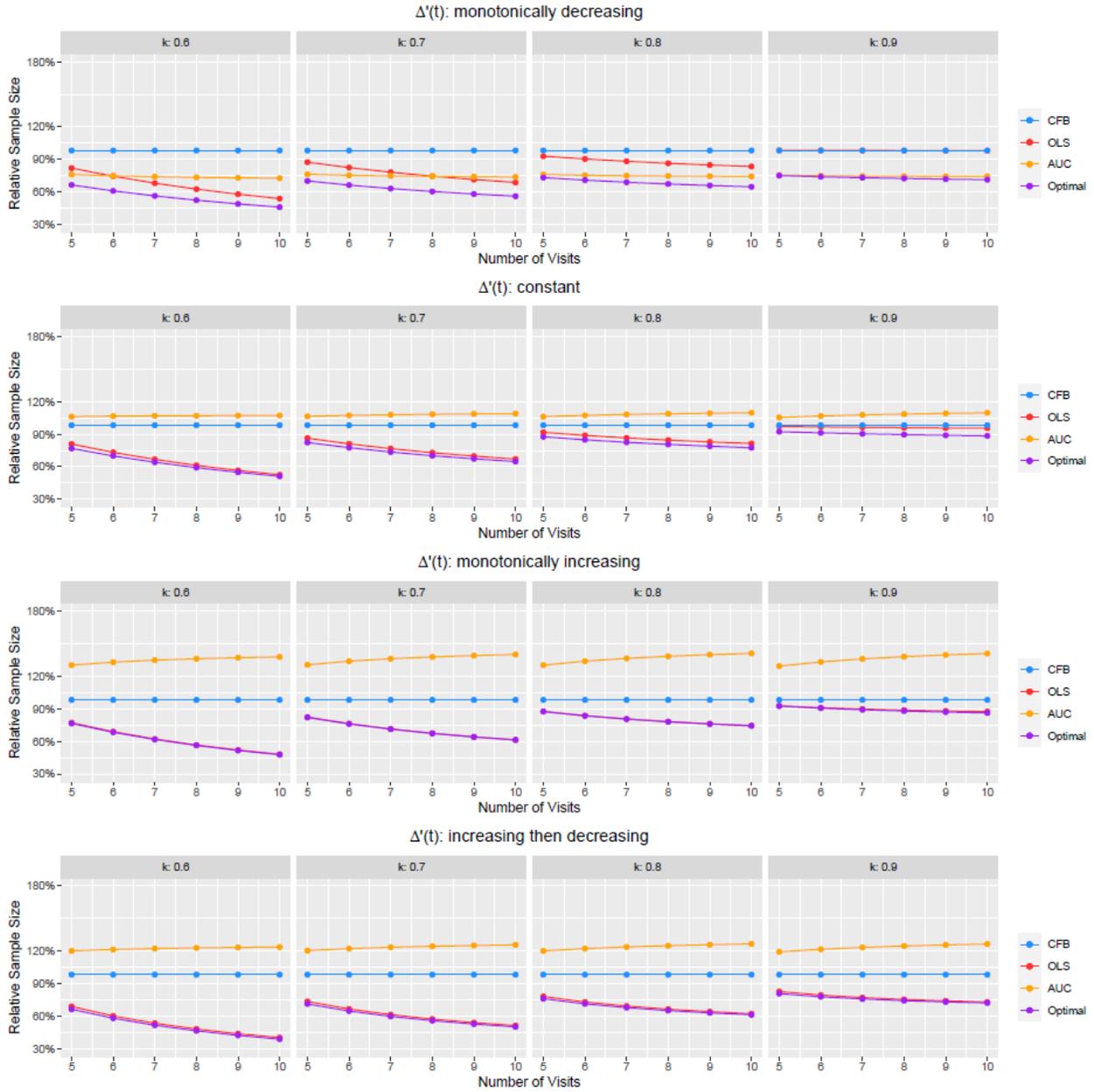

**Figure S6** Relative sample size for discrete PPRs. The CFB serves as the reference. Results based on "smart" coefficients $v_1^*$ and $\sigma = \sqrt{2}$.

## S6. Numerical results for the estimates based on continuous PPRs using the "smart" coefficients $q_1^*$

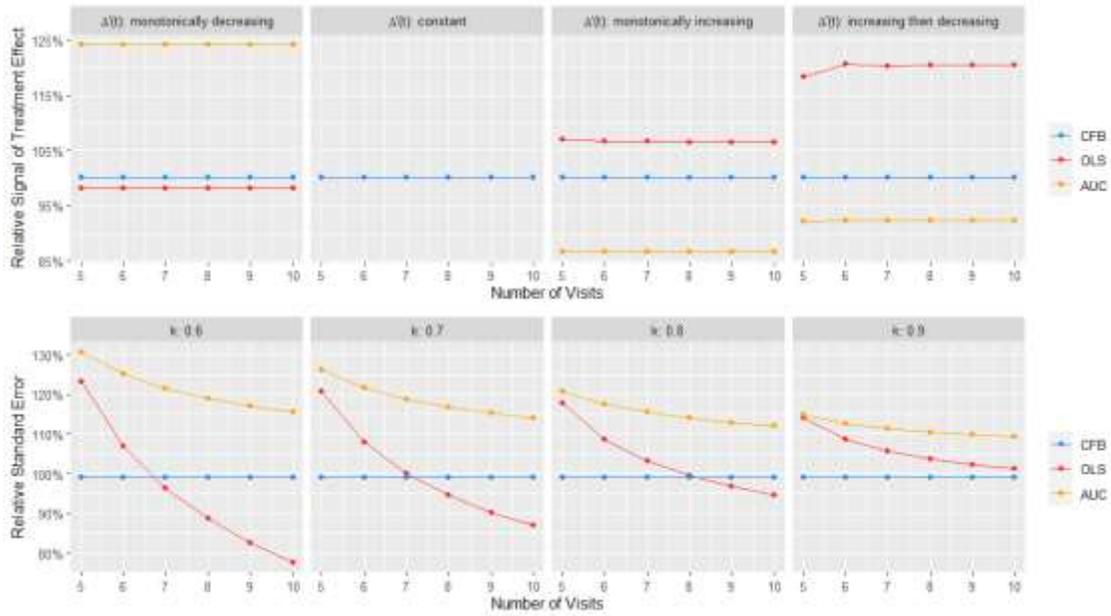

**Figure S7** Relative difference in signal of treatment effect (top panels) and standard error (bottom panels) for continuous PPRs. The CFB serves as the reference. Results based on "smart" coefficients $q_1^*$ and $\sigma = \sqrt{2}$.

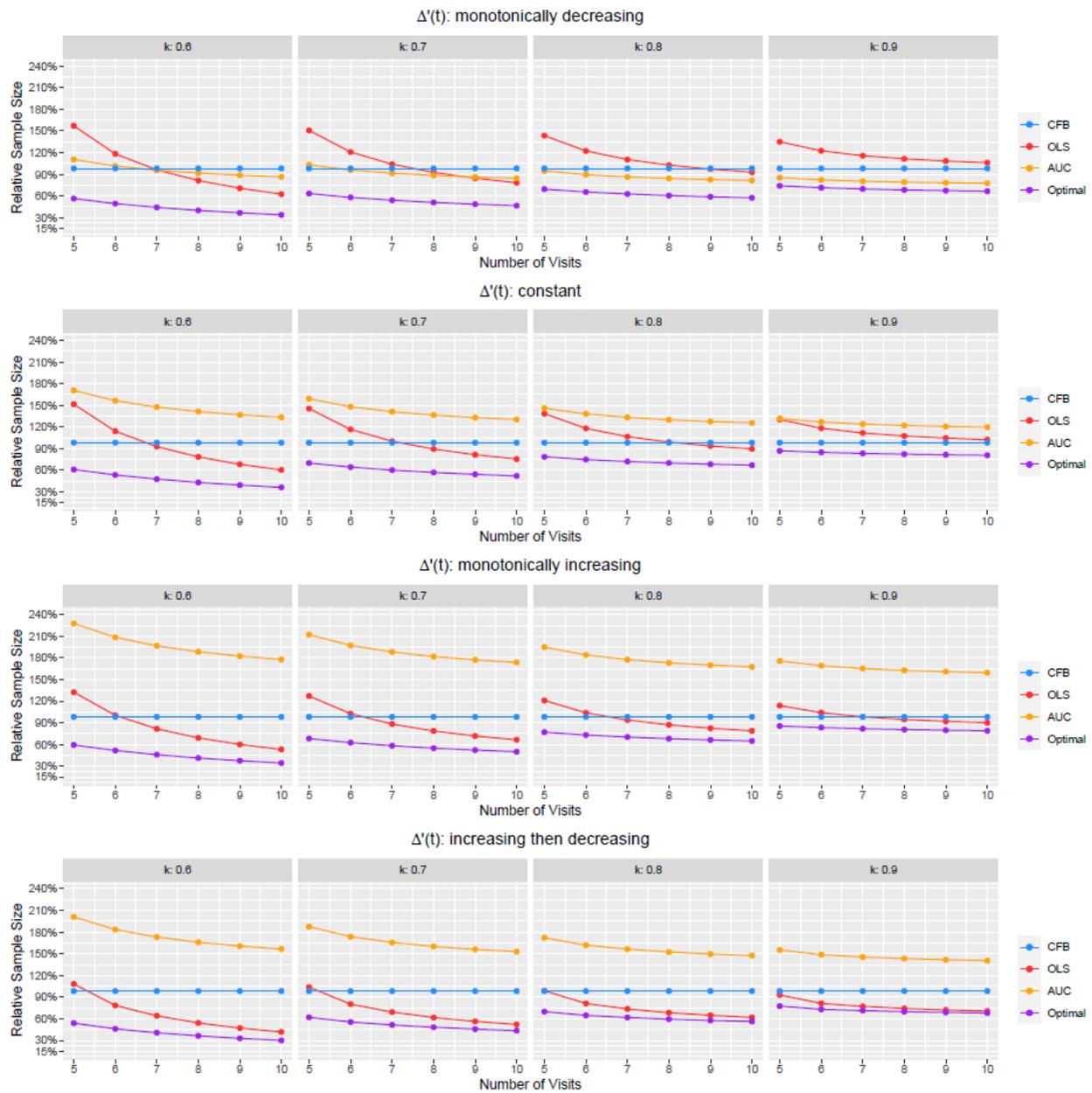

**Figure S8** Relative sample size for continuous PPRs. The CFB serves as the reference. Results based on "smart" coefficients $q_1^*$ and $\sigma = \sqrt{2}$.

**S7.** $w(t) = Beta(a, b)$ $(a > 1, b > 1) \to r_w(f)$ **is the slope of a weighted least square fit where the weight is *Beta(a-1,b-1)***

$$r_w(f) = \int_0^1 w(t) f'(t) dt = w(1) f(1) - w(0) f(0) - \int_0^1 w'(t) f(t) dt.$$

Because $w(0) = w(1) = 0$ for $w(t) = Beta(a, b)$, we have

$$r_w(f) = -\frac{\Gamma(a+b)}{\Gamma(a)\Gamma(b)} \int_0^1 [(a-1) t^{a-2} (1-t)^{b-1} - (b-1) t^{a-1} (1-t)^{b-2}] f(t) dt$$

$$= \frac{\Gamma(a+b)}{\Gamma(a)\Gamma(b)} \int_0^1 t^{a-2} (1-t)^{b-2} [(a+b-2) t - (a-1)] f(t) dt$$

$$= \frac{\int_0^1 \frac{\Gamma(a+b-2)}{\Gamma(a-1)\Gamma(b-1)} t^{a-2} (1-t)^{b-2} \left[t - \frac{a-1}{a+b-2}\right] f(t) dt}{\frac{(a-1)(b-1)}{(a+b-1)(a+b-2)^2}}$$

$$= \frac{\int_0^1 Beta(a-1, b-1) [t - mean(Beta(a-1, b-1))] f(t) dt}{\int_0^1 Beta(a-1, b-1) [t - mean(Beta(a-1, b-1))]^2 dt}.$$

Therefore, $r_w(f)$ is the slope of a weight least-square fit of $f(t)$ with the weight being $Beta(a-1, b-1)$.